\documentclass[aps,showpacs,preprintnumbers,amsmath,amssymb]{revtex4}
\UseRawInputEncoding
\usepackage{dcolumn}
\usepackage[dvips]{epsfig}
\usepackage{graphicx}
\usepackage{amssymb}
\usepackage{color}
\usepackage{enumerate}
\usepackage{multirow}

\begin{document}
\baselineskip=0.8 cm
\title{\bf Constraint on parameters of a rotating black hole in Einstein-bumblebee theory
by quasi-periodic oscillations}

\author{Zejun Wang$^{1}$,  Songbai Chen$^{1,2}$\footnote{Corresponding author: csb3752@hunnu.edu.cn},
Jiliang Jing$^{1,2}$ \footnote{jljing@hunnu.edu.cn}}
\affiliation{ $ ^1$ Department of Physics, Key Laboratory of Low Dimensional Quantum Structures
and Quantum Control of Ministry of Education, Synergetic Innovation Center for Quantum Effects and Applications, Hunan
Normal University,  Changsha, Hunan 410081, People's Republic of China
\\
$ ^2$Center for Gravitation and Cosmology, College of Physical Science and Technology, Yangzhou University, Yangzhou 225009, People's Republic of China}

\begin{abstract}
\baselineskip=0.6 cm
\begin{center}
{\bf Abstract}
\end{center}

We have studied quasi-periodic oscillations frequencies in a rotating black hole with Lorentz symmetry
breaking parameter in Einstein-bumblebee gravity by relativistic precession model. We find
that in the rotating case with non-zero spin parameter both of the periastron and nodal precession frequencies
increase with the Lorentz symmetry breaking parameter, but the azimuthal frequency decreases. In
the non-rotating black hole case, the nodal precession frequency disappears for arbitrary Lorentz
symmetry breaking parameter. With the observation data of GRO J1655-40, XTE J1550-564, and GRS 1915+105,  we find that the constraint on the Lorentz symmetry breaking parameter is more precise with data of GRO J1655-40 in which the best-fit value of the Lorentz symmetry breaking parameter is negative. This could lead to that the rotating black hole in Einstein-bumblebee gravity owns the higher Hawking temperature and the stronger Hawking radiation, but the lower possibility of exacting energy by Penrose process. However, in the range of $1 \sigma$, we also find that general relativity remains to be consistent with the observation data of GRO J1655-40, XTE J1550-564 and GRS 1915+105.
\end{abstract}

 \maketitle
\newpage
\section{Introduction}

Lorentz invariance has been of great importance in general relativity and the
standard model of particle physics. However, according to the development of unified gauge theories and the
signals from high energy cosmic rays \cite{lv01,lv02},  Lorentz symmetry may spontaneously break in the more fundamental physics at a higher scale of energy. And then studying Lorentz violation is also expected to obtain a deeper understanding of nature. In general, the direct test of Lorentz violation is impossible  because their high energy scale is unavailable in the current experimentations. However, recent investigations also show that some signals related to Lorentz violation could
emerge at lower energy scales so that their corresponding effects could be observed in
experiments \cite{lvbh1}.

Einstein-bumblebee gravity \cite{lvbh2} is a simple effective theory of gravity with Lorentz violation where the spontaneous breaking of Lorentz symmetry is induced by a nonzero vacuum expectation value of bumblebee vector field
$B_{\mu}$ with a suitable potential. The black hole solutions in Einstein-bumblebee gravity and the corresponding effects of Lorentz violation have been extensively studied in the past years \cite{lvbh2s, lvbh2s1,lvbhh1,lvbhh2,lvbhh12,lvbhh3,lvbhh4,lvbhh5,lvbhh6,lvbhh7,lvbhh8}. R. Casana \textit{et al} firstly found an exact solution of a static neutral black hole,  and discussed its some classical tests \cite{lvbh1}. And then,
the gravitational lensing \cite{lvbh3}, the Hawking radiation \cite{lvbh4} and quasinormal
modes \cite{lvbh5} have been addressed in this black hole spacetime.  Moreover, other spherically symmetric black hole
solutions, containing global monopole \cite{lvbh6}, cosmological constant \cite{lvbh7}, or Einstein-Gauss-Bonnet
term \cite{lvbh8}, and the traversable wormhole solution in the framework of the bumblebee gravity theory \cite{lvbh9} have also been found.  The cosmological implications of bumblebee gravity model are further investigated in \cite{lvbh10} . Furthermore,
the rotating black hole solution \cite{lvbhrot1} is also obtained in Einstein bumblebee
gravity, and the corresponding shadow \cite{lvbhrot1,lvbhrot1s}, accretion disk \cite{lvbhrot2}, superradiant instability of black hole \cite{lvbhrot3} and particle's motion \cite{lvbhrot4} around the black hole are studied.  A Kerr-Sen-like black hole with a bumblebee field has also been investigated \cite{lvbhrot5}.  These investigations are useful for testing Einstein bumblebee theory and detecting the effects caused by the Lorentz symmetry breaking originating from bumblebee vector field.

Quasi-periodic oscillations can be regarded as a promising arena to test the nature of the compact objects, which appear as peaks in the  observed X-ray power density spectrum emitted by accreting black hole binary systems \cite{qpo1,qpo2} and hold important information about
gravity in the strong field region.
Generally, the frequency range of the quasi-periodic oscillations changes from mHz to hundreds of Hz.  There are various theoretical models proposed to account for such peaks in power density spectrum, but the essence of quasi-periodic oscillations is still unclear at present. The relativistic precession model is a highly regarded model of explaining quasi-periodic oscillations in which  the oscillation frequencies are believed to associate with three fundamental frequencies of a test particle around a central object \cite{RPM1,RPM2,RPM20,RPM3,RPM301,RPM302}. In this model,  the azimuthal frequency $\nu_{\phi}$ and the periastron precession frequency $\nu_{\text{per}}$ of the test particle are explained, respectively, as the twin higher frequencies quasi-periodic oscillations. And  the nodal precession frequency $\nu_{\text{nod}}$ of the particle is identified with the low-frequency quasi-periodic peak in the power density spectrum of low-mass X-ray binaries. Thus, the low-frequency quasi-periodic signal is assumed to be emitted at the same orbit of the test particle where the twin higher frequencies signals are generated.
Together with the observation data of GRO J1655-40 \cite{RPM1}, the constraint on the black hole parameters in various theories of gravity  have been performed by quasi-periodic oscillations within the relativistic precession model \cite{TB1,TB101,TB2,TB3,TB4,TB5,TB6,TB7,TB8,TB9,TB10,TB11,TB12,TB13}. The main purpose of this paper is to constrain the Lorentz symmetry breaking parameter for a rotating black hole in Einstein-bumblebee theory of gravity by using of quasi-periodic oscillations with the observation data from GRO J1655-40,  XTE J1550-564 and GRS 1915+105 \cite{RPM1,TB13, XTE1,GRS1}.

The paper is organized as follows: In Sec.II, we will review briefly the rotating black hole in Einstein-bumblebee theory of gravity \cite{lvbhrot1}. In Sec.III, we study quasi-periodic oscillations in the above black hole spacetime and then
make a constraint on the  Lorentz symmetry breaking parameter  with the observation data of GRO J1655-40,  XTE J1550-564 and GRS 1915+105. Finally, we present a summary.

\section{A rotating black hole in Einstein-bumblebee theory of gravity}

In this section we review briefly a  rotating black hole in Einstein-bumblebee theory \cite{lvbhrot1}. In the framework of the bumblebee gravity theory, the spontaneous Lorentz symmetry breaking is induced by a vector $B_{\mu}$ with a non-zero nonzero vacuum expectation value.  Through a coupling, the bumblebee vector field $B_{\mu}$ would affect the dynamics of the gravitational field. The action describing such kind of Lorentz symmetry breaking is \cite{lvbh1,lvbh2, lvbh2s,lvbh2s1}
\begin{eqnarray}\label{action}
S=\int d^4\sqrt{-g}\bigg[\frac{1}{16\pi}(R+\xi B^{\mu\nu}R_{\mu\nu})-\frac{1}{4}B^{\mu\nu}B_{\mu\nu}-V(B_{\mu}B^{\mu}\pm b^2)\bigg],
\end{eqnarray}
where $\xi$ is the coupling constant with the dimension $M^{-1}$  and the bumblebee field strength $B_{\mu\nu}=\partial_{\mu}B_{\nu}-\partial_{\nu}B_{\mu}$. The potential $V$, inducing Lorentz violation, has a minimum at $B_{\mu}B^{\mu}\pm b^2=0$ ( where $b$ is a real positive constant), which drives a nonzero vacuum value $\langle B_{\mu}\rangle=b_{\mu}$ with $b_{\mu}b^{\mu}=\mp b^2$. The  signs ``$\pm$" in the potential  determine whether the field $b_{\mu}$ is timelike or spacelike. Then the nonzero
vector background $b_{\mu}$ spontaneously breaks the Lorentz
symmetry \cite{lvbh1,lvbh2, lvbh2s,lvbh2s1}.
The extended vacuum Einstein equations in this model with Lorentz symmetry breaking  becomes
\begin{eqnarray}\label{dmmass}
R_{\mu\nu}-\frac{1}{2}g_{\mu\nu}R=T_{\mu\nu},
\end{eqnarray}
with
\begin{eqnarray}
T_{\mu\nu}&=&B_{\mu\alpha}B_{\;\nu}^{\alpha}-g_{\mu\nu}\bigg(\frac{1}{4}B_{\alpha\beta}B^{\alpha\beta}+V\bigg)-2B_{\mu}B_{\nu}V'
+\frac{\xi}{8\pi}\bigg[\frac{1}{2}g_{\mu\nu}B_{\alpha}B^{\alpha}-B_{\mu}B^{\alpha}R_{\alpha\nu}-B_{\nu}B^{\alpha}R_{\alpha\mu}
\nonumber\\
&+&\frac{1}{2}\nabla_{\alpha}\nabla_{\mu}(B^{\alpha} B_{\nu})+\frac{1}{2}\nabla_{\alpha}\nabla_{\nu}(B^{\alpha} B_{\mu})-\frac{1}{2}\nabla^2(B_{\mu} B_{\nu})-\frac{1}{2}g_{\mu\nu}\nabla_{\alpha}\nabla_{\beta}(B^{\alpha} B_{\beta})\bigg].
\end{eqnarray}
The Einstein equations (\ref{dmmass}) admits a rotating black hole solution with a metric \cite{lvbhrot1}
\begin{eqnarray}\label{metric}
ds^{2}&=&-\bigg(1-\frac{2Mr}{\rho^2}\bigg)dt^{2}-
\frac{4Mar\sqrt{l+1}\sin^2\theta}{\rho^2}dtd\phi
+\frac{\rho^2}{\Delta} dr^{2}+\rho^2 d\theta^{2}\nonumber\\ &+&\frac{\sin^2\theta}{\rho^2}\bigg[\bigg(r^2+(l+1)a^2\bigg)^2-\Delta (l+1)^2a^2\sin^2\theta \bigg]d\phi^2,
\end{eqnarray}
where
\begin{eqnarray}
\rho^2=r^2+(l+1)a^2\cos{\theta}^2,\quad\quad\quad\Delta=\frac{r^{2}-2Mr}{l+1}+a^{2}.
\end{eqnarray}
Here $M$ is the ADM mass and $a$ is the spin parameter of black hole. The form of the bumblebee field is $b_{\mu}=(0,b\rho, 0, 0)$,  and the parameter $l=\xi b^2$ depends on the spontaneous Lorentz symmetry breaking of the vacuum of the Einstein-bumblebee vector field.  The determinant of the metric (\ref{metric}) is $g=-(l+1)\rho^4\sin^2\theta $ and then the metric becomes degenerate when $l=-1$. Thus, in order to maintain its Lorentz signature, one must have $l>-1$, which means that the coupling $\xi$ should be restricted to $\xi>-\frac{1}{b^2}$. As in the Kerr black hole case, the singularity lies at $\rho^2=0$ and the horizon locates at $\Delta=0$. However, the horizon radius becomes
\begin{eqnarray}
r_{\pm}=M\pm\sqrt{M^2-(l+1)a^2},
\end{eqnarray}
which depends on the spontaneous Lorentz symmetry breaking parameter $l$. With the increase of the absolute value of $l$,
the outer horizon radius decreases for the positive $l$ and increases for the negative one. Thus, comparing with the usual Kerr black hole, the negative $l$ leads to that the rotating black hole (\ref{metric}) owns the higher Hawking temperature and the stronger Hawking radiation \cite{lvbhrot1} . Moreover, for a rotating black hole (\ref{metric}), its mass and spin parameters must satisfy $\frac{|a|}{M}\leq\frac{1}{\sqrt{l+1}}$. The negative $l$ broadens the range of black hole spin parameter so that $|a|>M$, but the positive $l$ shortens the
range of $a$, which differs quite from the Kerr case in general relativity.

\section{Constraint on parameters of a rotating black hole in
Einstein-bumblebee theory by quasi-periodic oscillations}

\begin{figure}
\includegraphics[width=6cm ]{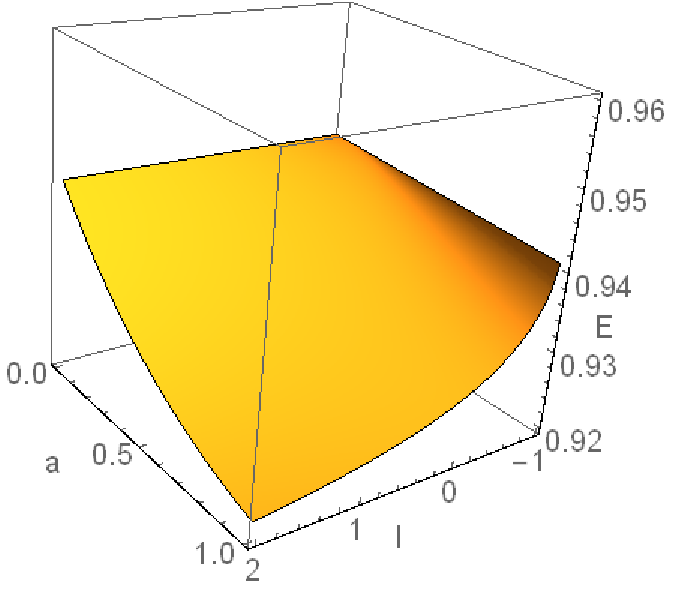}
\caption{ The equivalent surface of the circular orbit radius $r_0=6.5$ in the  parameter space ($a$,  $l$, $E$)  in the rotating black hole spacetime (\ref{metric}) in Einstein-bumblebee theory. Here we set $M=1$.}
\label{r0ps}
\end{figure}

In this section, we will apply quasi-periodic oscillations to make a constraint on parameters of a rotating black hole (\ref{metric}) in Einstein-bumblebee theory.  For a general stationary and axially symmetric spacetime, the metric of a rotating black hole with bumblebee field (\ref{metric}) can be written as a common form
\begin{eqnarray}
ds^2&=&g_{tt}dt^2+g_{rr}dr^2+2g_{t\phi}dtd\phi+g_{\theta\theta}d\theta^2
+g_{\phi\phi}d\phi^2. \label{metric3n}
\end{eqnarray}
Obviously, the metric coefficients in Eq. (\ref{metric}) are independent of the coordinates $ t$ and $\phi$. Thus, the geodesic motion of particle in the black hole spacetime (\ref{metric}) exists two conserved quantities, i.e., the specific energy at infinity $E$ and the conserved $z$-component of the specific angular momentum at infinity $L_z$, and the forms of $E$ and $L_z$ can be expressed as
\begin{eqnarray}
E=-p_{t}=-g_{tt}\dot{t}-g_{t\phi}\dot{\phi}, \quad \quad \quad L_{z}=p_{\phi}=g_{t\phi}\dot{t}+g_{\phi\phi}\dot{\phi}.\label{conserved quantities}
\end{eqnarray}
With above two conserved quantities, the timelike geodesics can be further simplified as
\begin{eqnarray}
&&\dot{t}=\frac{g_{\phi\phi}E+g_{t\phi}L_z}{g^2_{t\phi}-g_{tt}g_{\phi\phi}},\label{u1}\\
&&\dot{\phi}=\frac{g_{t\phi}E+g_{tt}L_z}{g_{tt}g_{\phi\phi}-g^2_{t\phi}},\label{u2}\\
&&g_{rr}\dot{r}^2+g_{\theta\theta}\dot{\theta}^2=V_{eff}(r,\theta; E,L_z),\label{u3}
\end{eqnarray}
where $V_{eff}(r,\theta; E,L_z)$ is the effective potential with the form
\begin{eqnarray}\label{veffpo}
V_{eff}(r,\theta; E,L_z)=\frac{E^2g_{\phi\phi}+2EL_zg_{t\phi}+L^2_zg_{tt}
}{g^2_{t\phi}-g_{tt}g_{\phi\phi}}-1.
\end{eqnarray}
Here the overhead dot represents a derivative with respect to the
affine parameter $\lambda$. The effective potential determines the orbit of the particle.  The form of potential (\ref{veffpo}) in the equatorial plane becomes
\begin{eqnarray}\label{vpequator}
V_{eff}(r,\frac{\pi}{2}; E,L_z)=\frac{[r^3+(r+2M)(l+1)a^2]E^2-4aM\sqrt{l+1}EL_z-(r-2M)L^2_z
}{r[r^2-2Mr+(l+1)a^2]}-1.
\end{eqnarray}
Actually, the radial component of the timelike geodesic equations
\begin{eqnarray}
\frac{d}{d\lambda}(g_{\mu\nu}\dot{x}^{\nu})=\frac{1}{2}(\partial_{\mu}g_{\nu\rho})\dot{x}^{\nu}\dot{x}^{\rho},
\end{eqnarray}
can be written as \cite{RPM1,RPM2,RPM20,RPM3}
\begin{eqnarray}\label{cedx0r}
\frac{d}{d\lambda}(g_{rr}\dot{r})=
\frac{1}{2}\bigg[(\partial_{r}g_{tt})\dot{t}^2+2(\partial_{r}g_{t\phi})\dot{t}\dot{\phi}+(\partial_{r}g_{\phi\phi})\dot{\phi}^2+(\partial_{r}g_{rr})\dot{r}^2+(\partial_{r}g_{\theta\theta})\dot{\theta}^2\bigg].
\end{eqnarray}
We here consider only the case where a particle moving along a circular orbit in the equatorial plane, i.e., $r=r_0$ and $\theta=\pi/2$, which means that $\dot{r}=\dot{\theta}=\ddot{r}=0$.
Thus, for the circular equatorial orbit case,  Eq.(\ref{cedx0r}) can be simplified as
\begin{eqnarray}\label{tdot0}
(\partial_{r}g_{tt})\dot{t}^2+2(\partial_{r}g_{t\phi})\dot{t}\dot{\phi}+(\partial_{r}g_{\phi\phi})\dot{\phi}^2=0,
\end{eqnarray}
which gives the orbital angular velocity $\Omega_{\phi}$ of a particle moving along the circular orbits
\begin{eqnarray}
\Omega_{\phi}=\frac{d\phi}{dt}=\frac{-g_{t\phi,r}\pm\sqrt{(g_{t\phi,r})^2
+g_{tt,r}g_{\phi\phi,r}}}{g_{\phi\phi,r}}=\pm\frac {g_{tt,r}}{\sqrt{(g_{t\phi,r})^2
+g_{tt,r}g_{\phi\phi,r}}\pm g_{t\phi,r}},\label{jsd0}
\end{eqnarray}
here the sign  is $+ (-)$ for co-rotating (counter-rotating)
orbits. The corresponding azimuthal frequency $\nu_{\phi}=\Omega_{\phi}/(2\pi)$.
For a timelike particle moving along circular orbits in the equatorial plane, the timelike conditions
$g_{\mu\nu}\dot{x}^{\mu}\dot{x}^{\nu}=-1$ gives another relationship between $\dot{t}$ and $\dot{\phi}$
\begin{eqnarray}\label{tdot1}
g_{tt}\dot{t}^2+2g_{t\phi}\dot{t}\dot{\phi}+g_{\phi\phi}\dot{\phi}^2=-1.
\end{eqnarray}
From two independent equations (\ref{tdot0}) and (\ref{tdot1}), one can obtain
\begin{eqnarray}\label{t0}
\dot{t}=\frac{1}{\sqrt{-g_{tt}-2g_{t\phi}\Omega_{\phi}-g_{\phi\phi}\Omega_{\phi}^2}}.
\end{eqnarray}
Together with Eq.(\ref{conserved quantities}), one can find that the specific energy  $E$ and the conserved $z$-component of the specific angular momentum  $L_z$ are expressed respectively as \cite{RPM1,RPM2,RPM20,RPM3}
\begin{eqnarray}
&&E=-\frac{g_{tt}+g_{t\phi}\Omega_{\phi}}{\sqrt{-g_{tt}-2g_{t\phi}\Omega_{\phi}
-g_{\phi\phi}\Omega^2_{\phi}}},\nonumber\\
&&L_z=\frac{g_{t\phi}+g_{\phi\phi}\Omega_{\phi}}{\sqrt{-g_{tt}
-2g_{t\phi}\Omega_{\phi}-g_{\phi\phi}\Omega^2_{\phi}}}.\label{jsd}
\end{eqnarray}
The radius of circular orbit $r_0$ in the equatorial plane can be given by the conditions
\begin{eqnarray} \label{vcondition}
V_{eff}(r_0,\frac{\pi}{2}; E,L_z)=0,\quad\quad\quad \frac{d V_{eff}(r,\frac{\pi}{2}; E,L_z)}{dr}\bigg|_{r=r_0}=0.
\end{eqnarray}
Making use of these two conditions, we can obtain the specific angular momentum  $L_z$ of a particle moving along the circular orbit $r_0$ in the equatorial plane
\begin{eqnarray}\label{L0gdao}
L_z=\pm\sqrt{3(E^2-1)[r^2_0+(l+1)a^2]+4Mr_0},
\end{eqnarray}
and find that the corresponding circular orbit  $r_0$ satisfies
\begin{eqnarray}\label{r0gdao}
(1-E^2)r^3_0+M(3E^2-4)r^2_0+4M^2r_0+Ma^2(l+1)(2E^2-1)-2aEM\sqrt{(l+1)W}=0,
\end{eqnarray}
with
\begin{eqnarray}
W=3(E^2-1)r^2_0+4Mr_0+a^2(E^2-1)(l+1).
\end{eqnarray}
It indicates the radius of circular orbit $r_0$ is a function of four independent parameters, i.e., $M$, $a$, $l$ and the particle's energy $E$. Thus, the circular orbit with certain fixed radius $r_0$ could exists for a particle in a rotating black hole spacetime (\ref{metric}) in Einstein-bumblebee theory since there are four adjustable parameters. In Fig. (\ref{r0ps}), we present the equivalent surface of the circular orbit radius $r_0=6.5$ in the parameter space $a$, $l$ and $E$ ( here we set $M=1$), which shows that it is possible for the existence of circular orbit with $r_0=6.5$  for fixed $l$ and $a$ through the choice of a proper parameter $E$. For the non-rotating black hole (i.e., $a=0$), we find that
\begin{eqnarray}
r_0=\frac{[(3E^2-4)\pm E\sqrt{9E^2-8}]M}{2(E^2-1)},
\end{eqnarray}
which is independent of the parameter $l$. This can be explained by a fact that the potential (\ref{vpequator}) does not depend on $l$ as $a=0$. The radius $r_0$ has positive roots as $E\geq\frac{2\sqrt{2}}{3}$ and no any real root as $E<\frac{2\sqrt{2}}{3}$. These positive roots increase with the black hole mass $M$. With the increase of $E$, the root with the sign ``+" decreases in the allowable range of $E$, but the root with the sign ``$-$" increases as $\frac{2\sqrt{2}}{3}\leq E<1$ and it becomes negative as $E>1$. For the rotating case with $a\neq0$, we can not obtain the analytical form of $r_0$. From Eq.(\ref{r0gdao}), we can get the partial derivative of $r_0$ with respect to $M$, $a$, $l$ and $E$, respectively.
\begin{eqnarray}
&&\frac{\partial r_0}{\partial M}\bigg|_{a,l,E}=\frac{[(3E^2-4)r^2_0+8Mr_0+a^2(l+1)(2E^2-1)]\sqrt{W}-2 aE \sqrt{l+1}(2Mr_0+W)}{[3(E^2-1)r_0+2M][(r_0-2M)\sqrt{W}+2aME\sqrt{l+1}]},\\
&&\frac{\partial r_0}{\partial a}\bigg|_{M,l,E}=-\frac{2M\sqrt{l+1}[EW-(2E^2-1)a\sqrt{W(l+1)}+ a^2E(E^2-1)(l+1)] }{[3(E^2-1)r_0+2M][(r_0-2M)\sqrt{W}+2aME\sqrt{l+1}]},\label{dsh2}\\
&&\frac{\partial r_0}{\partial l}\bigg|_{M,a,E}=-\frac{aM[EW-(2E^2-1)a\sqrt{W(l+1)}+ a^2E(E^2-1)(l+1)] }{\sqrt{l+1}[3(E^2-1)r_0+2M][(r_0-2M)\sqrt{W}+2aME\sqrt{l+1}]},\label{dsh3}\\
&&\frac{\partial r_0}{\partial E}\bigg|_{M,a,l}=-\frac{2aM\sqrt{l+1}[3E^2r^2_0+a^2E^2(l+1)+W-2aE\sqrt{W(l+1)}]+2Er^2_0(r_0-3M)\sqrt{W} }{[3(E^2-1)r_0+2M][(r_0-2M)\sqrt{W}+2aME\sqrt{l+1}]}.
\end{eqnarray}
The above formulas indicate that it is not easy to determine the signs of these partial derivative determine, which means that  the change of circular orbital radius $r_0$  with  $M$, $a$, $l$ and $E$ becomes very complicated in the rotating black hole case. However,  formulas (\ref{dsh2}) and (\ref{dsh3}) tell us that the dependent behavior of $r_0$ on the parameter $l$ is qualitatively similar to that on the spin parameter $a$.
\begin{figure}
\includegraphics[width=5cm ]{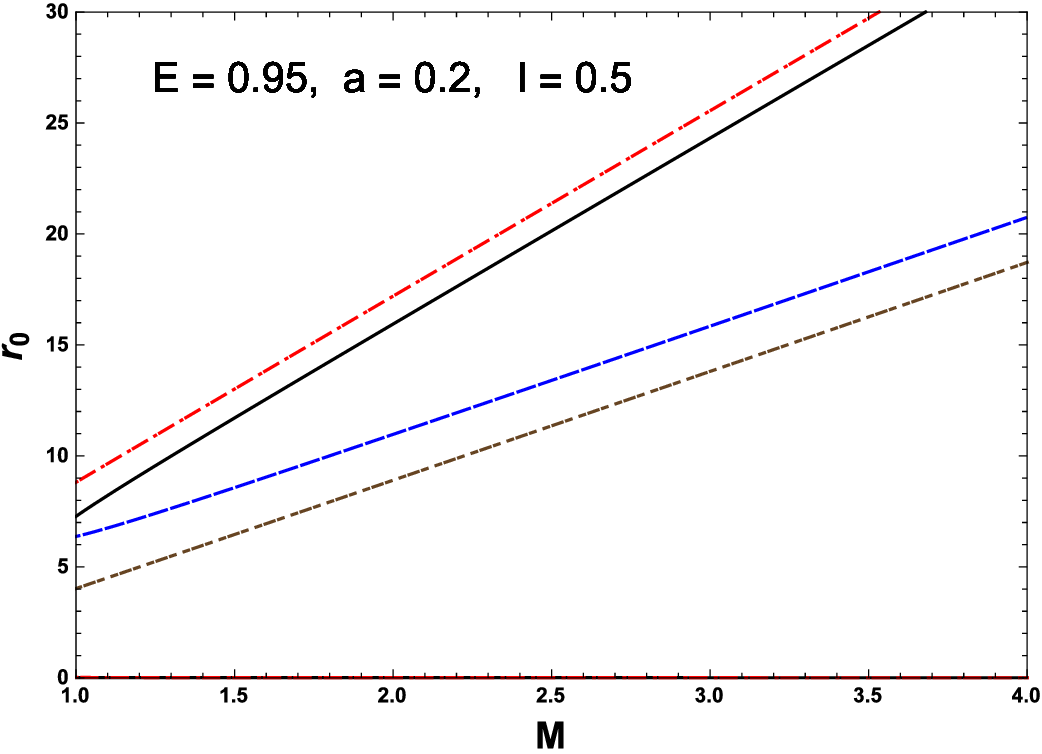}\includegraphics[width=5cm ]{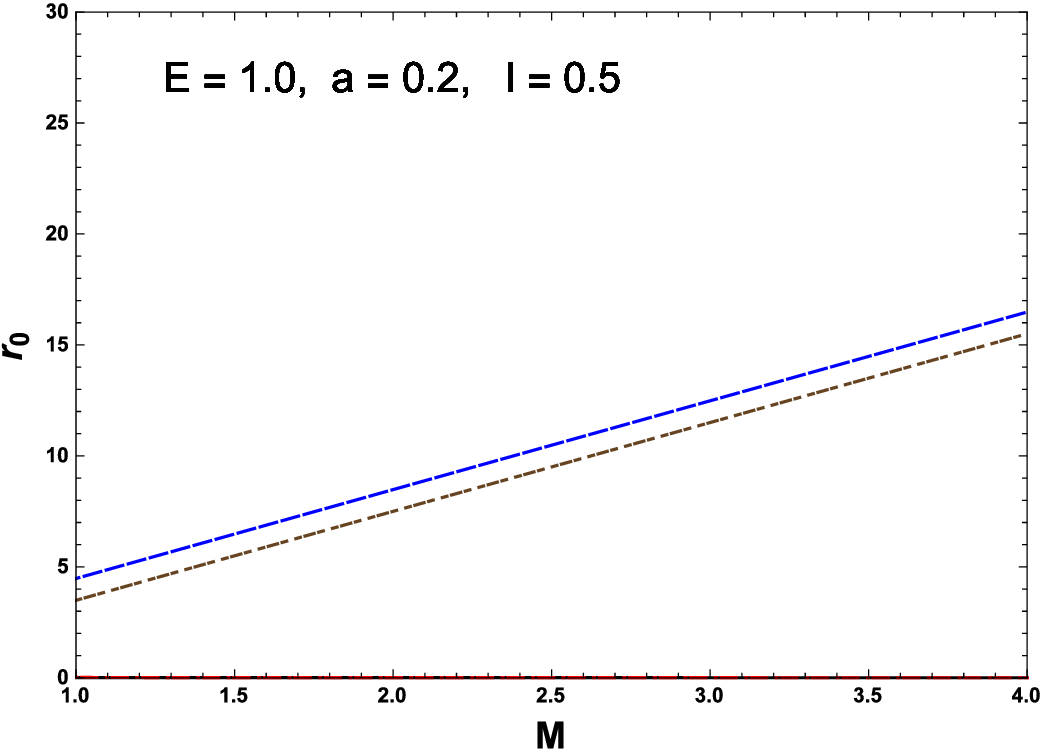}\includegraphics[width=5cm ]{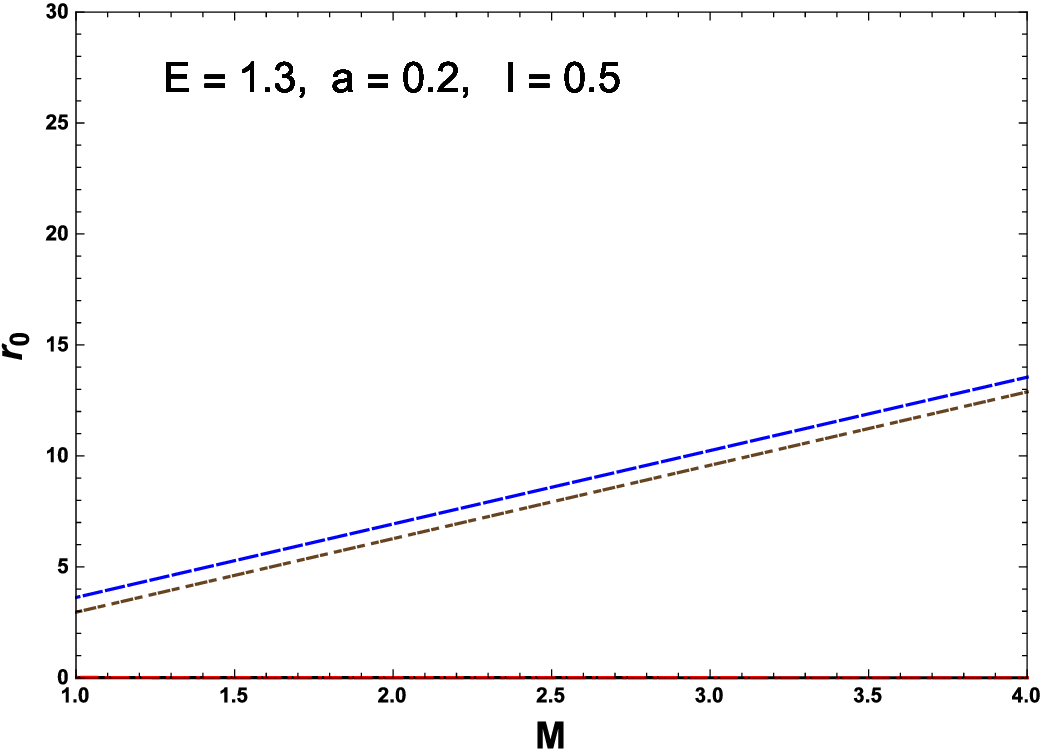}\\
\includegraphics[width=5cm ]{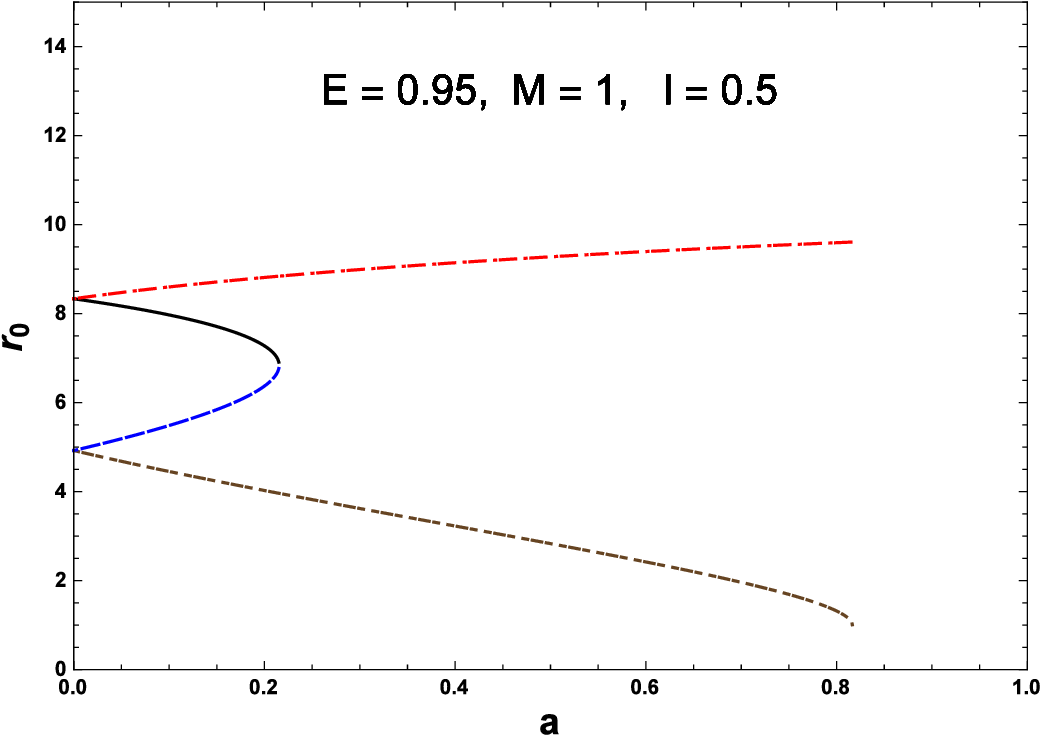}\includegraphics[width=5cm ]{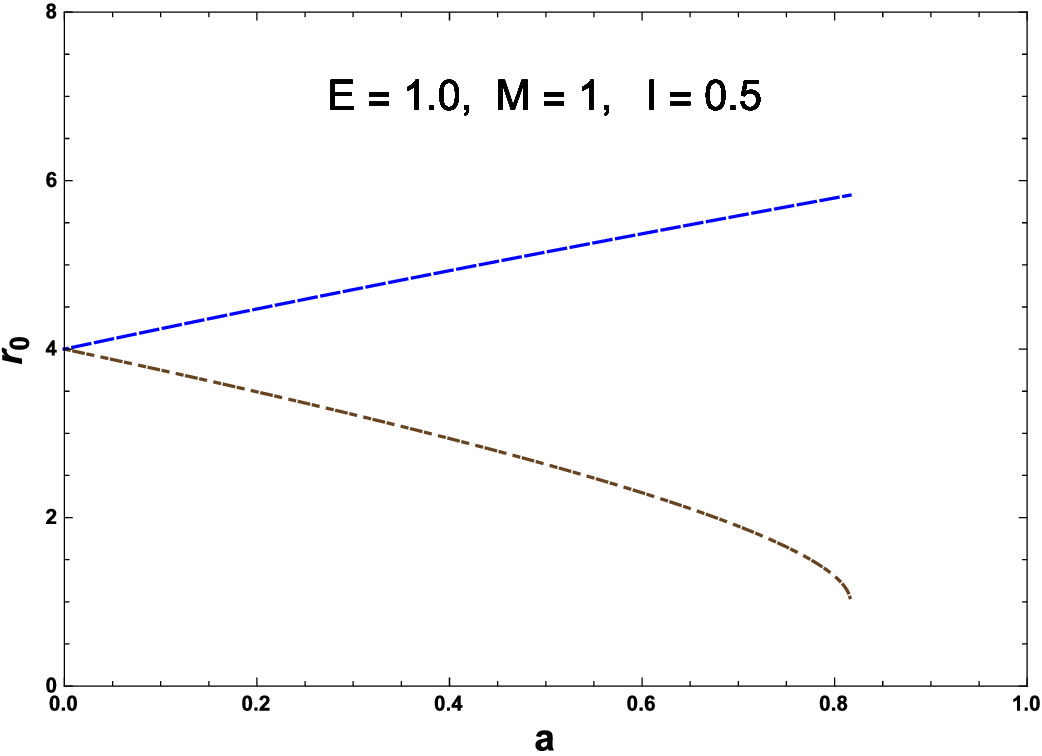}\includegraphics[width=5cm ]{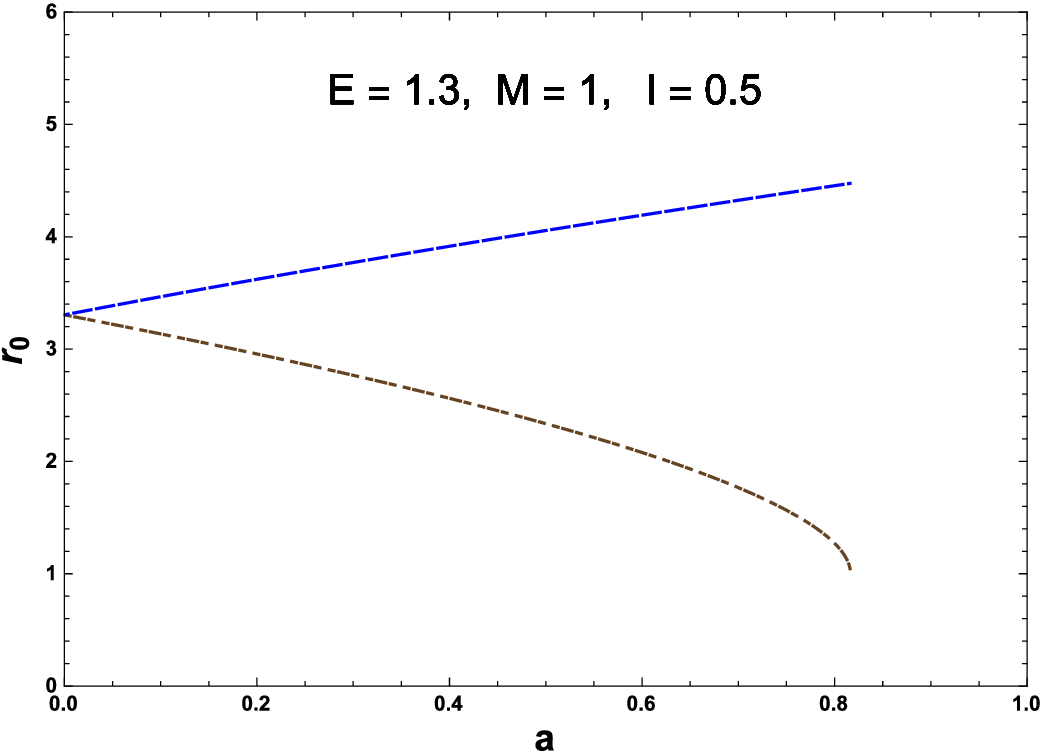}\\
\includegraphics[width=5cm ]{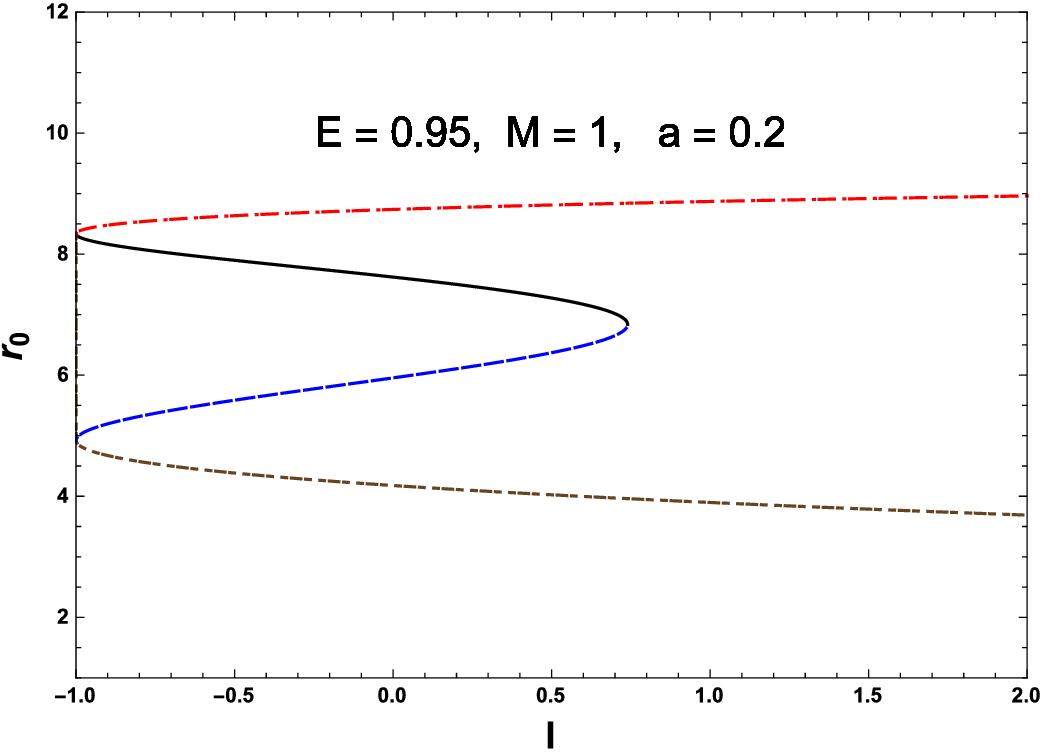}\includegraphics[width=5cm ]{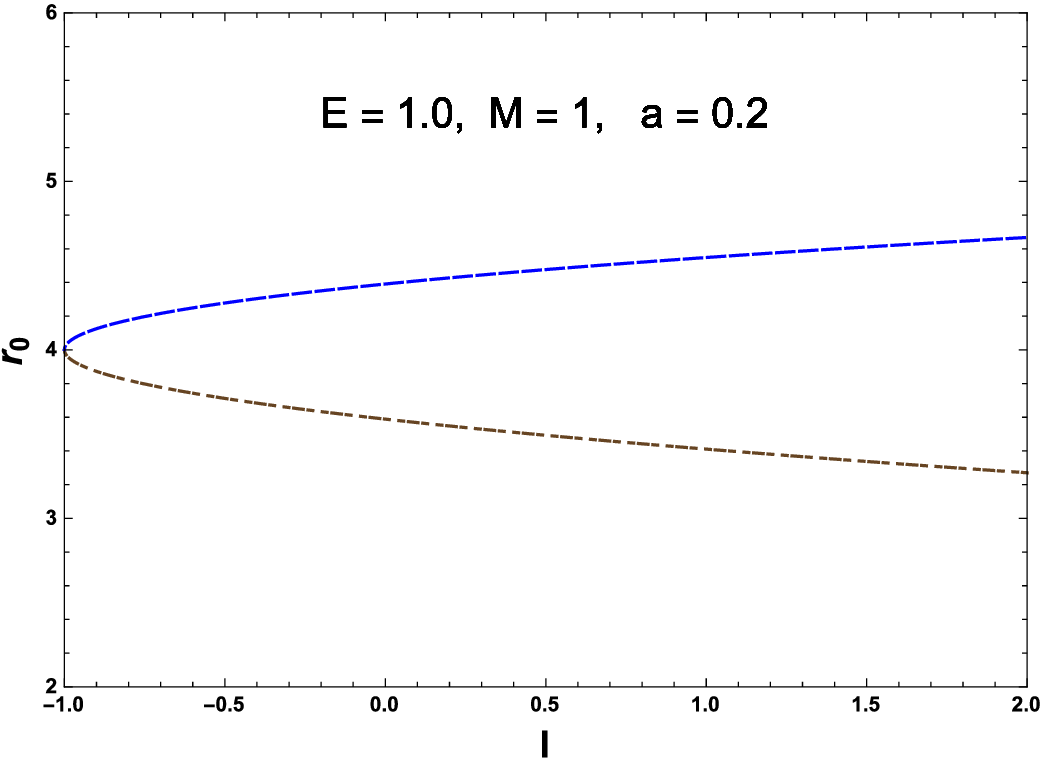}\includegraphics[width=5cm ]{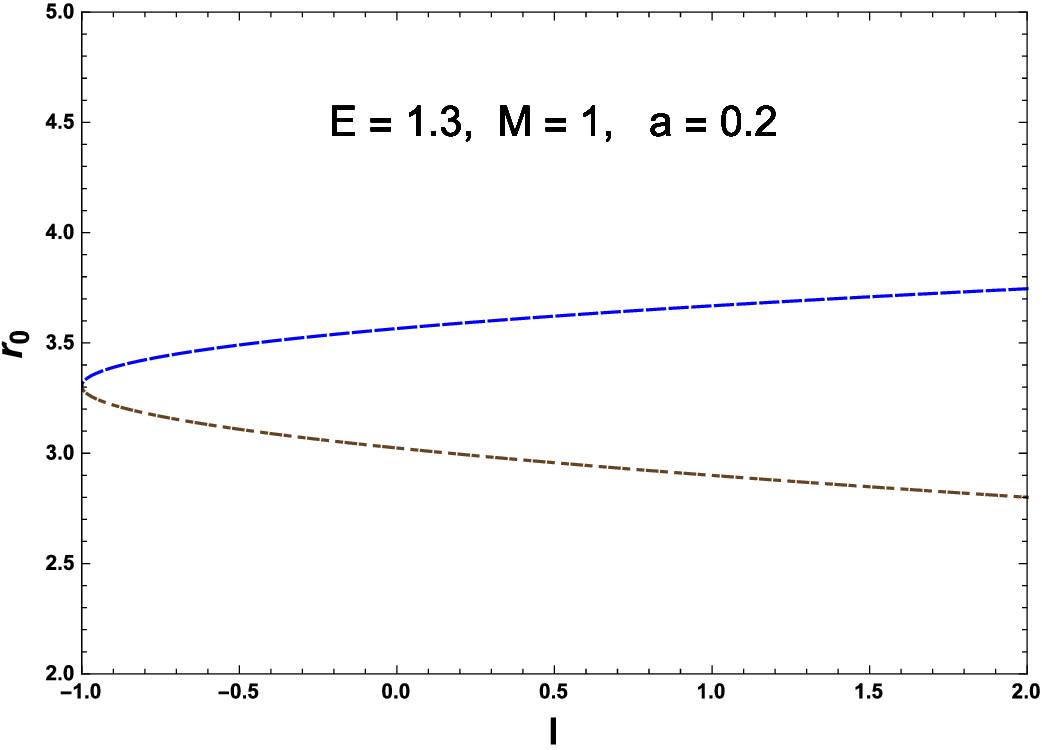}\\
\includegraphics[width=5cm ]{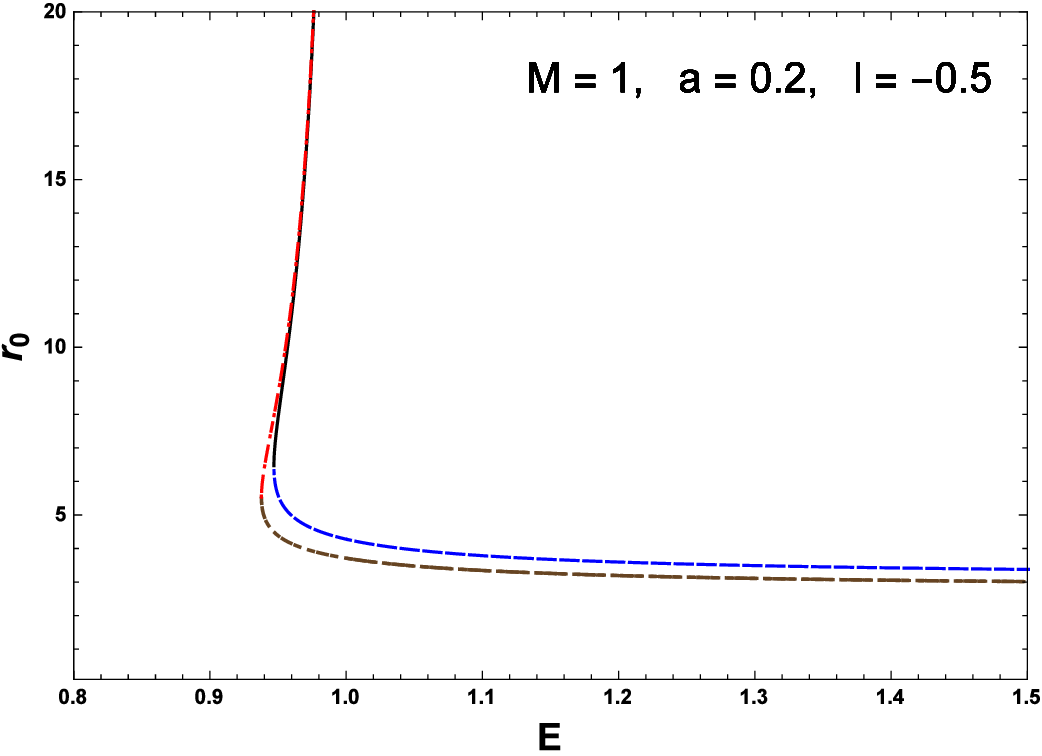}\includegraphics[width=5cm ]{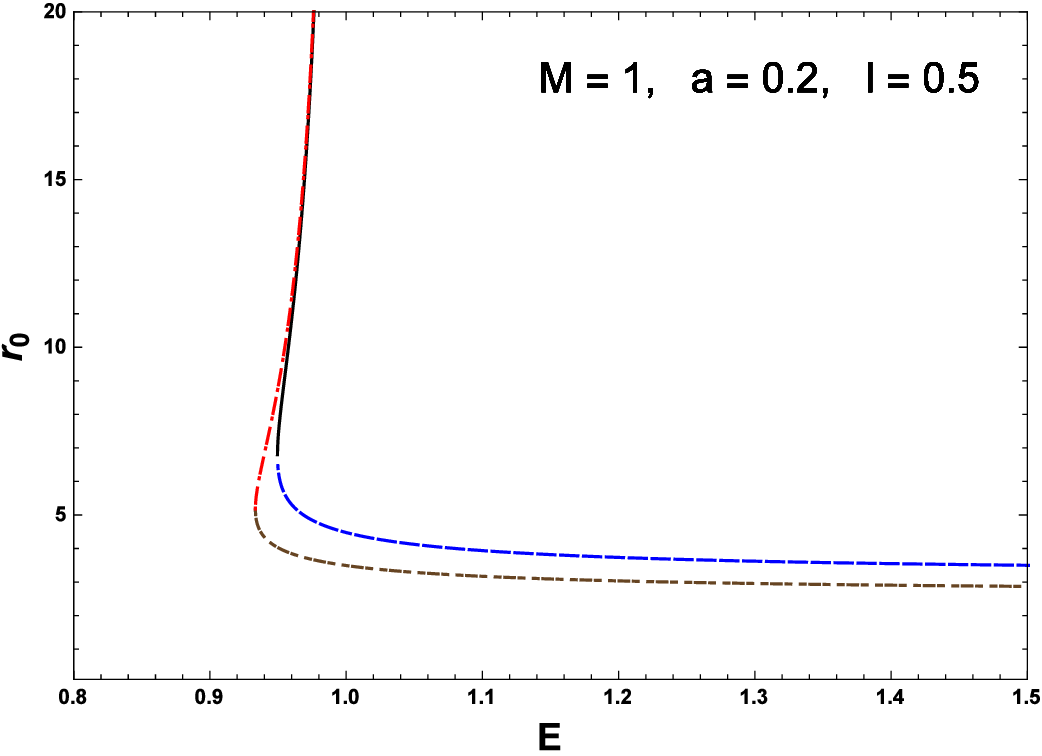}\includegraphics[width=5cm ]{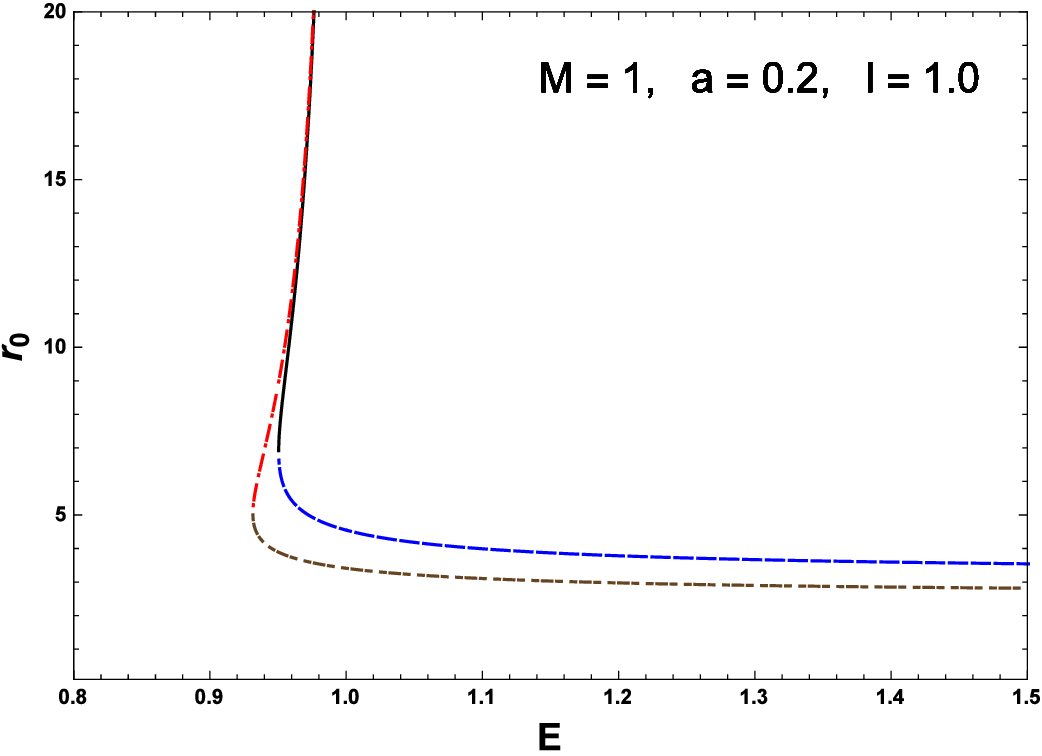}
\caption{The change of the  circular orbit radius $r_0$ with the black hole parameters $M$, $a$, $l$ and the particle's energy $E$ in the rotating black hole spacetime (\ref{metric}) in Einstein-bumblebee theory. In each panel, the red or black line denotes the unstable orbit, the blue or brown line corresponds to the stable orbit. }
\label{radius0}
\end{figure}
In Fig.(\ref{radius0}), we present the change of circular orbital radius $r_0$  with  $M$, $a$, $l$ and $E$ for some fixed parameters. For the chosen parameters, as the particle's energy $E<1$, there exist four circular orbits: a stable co-rotating orbit,  a stable counter-rotating orbit,  an unstable counter-rotating orbit
and an unstable co-rotating orbit, which are marked in the brown, blue, black  and red lines, respectively.
 While as $E\geq1$, there are two circular orbits, which correspond to the stable co-rotating orbit and the stable counter-rotating one, respectively. With the increase of  black hole mass parameter $M$, the radius $r_0$ for each circular orbit is an increasing function of black hole mass parameter $M$ as $a=0.2$ and $l=0.5$. With the increasing spin parameter $a$,  the radius $r_0$ for the unstable co-rotating orbit and the stable counter-rotating orbit increases, but deceases for the another two orbits. As in the previous discussion, Fig.(\ref{radius0}) also shows that the change of $r_0$ with the parameter $l$ is similar to that with $a$.  From Fig.(\ref{radius0}), as $E\geq1$, we find that the radius $r_0$ for both of circular orbits decreases with $E$. However, as $E<1$,  the radius $r_0$ for two stable circular orbits decrease with $E$, but increases for another two unstable orbits.

\begin{figure}
\includegraphics[width=5.5cm ]{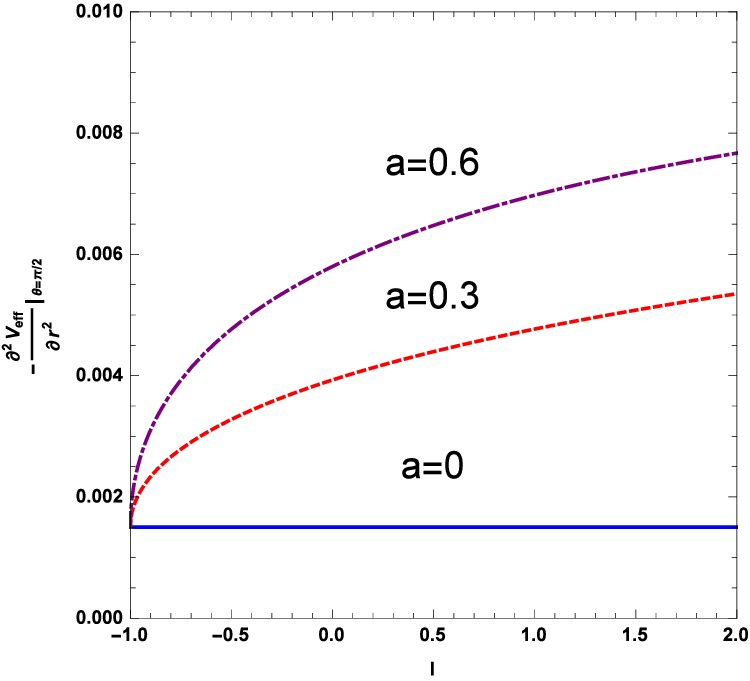} \includegraphics[width=5.5cm ]{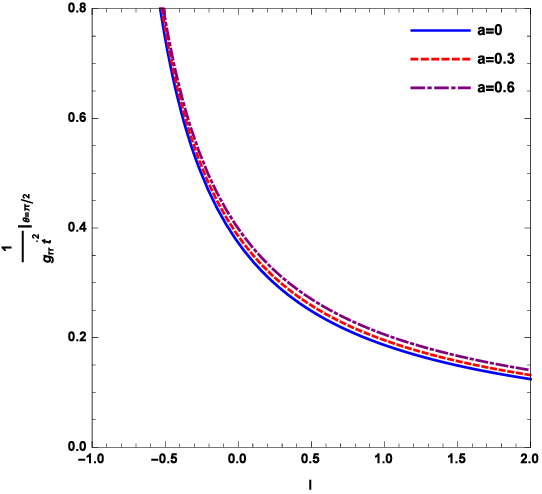}\includegraphics[width=5.2cm ]{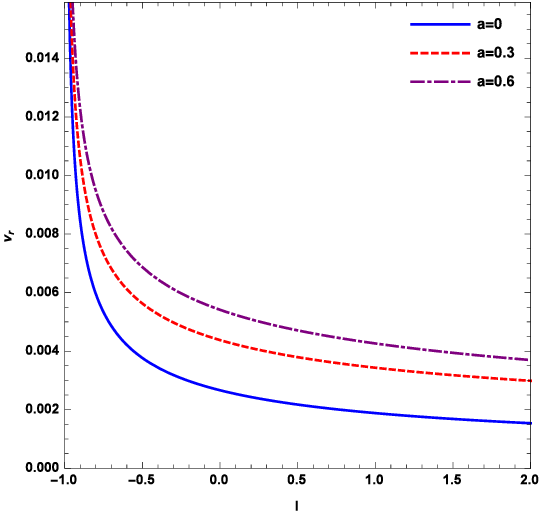}\\
\includegraphics[width=5.2cm ]{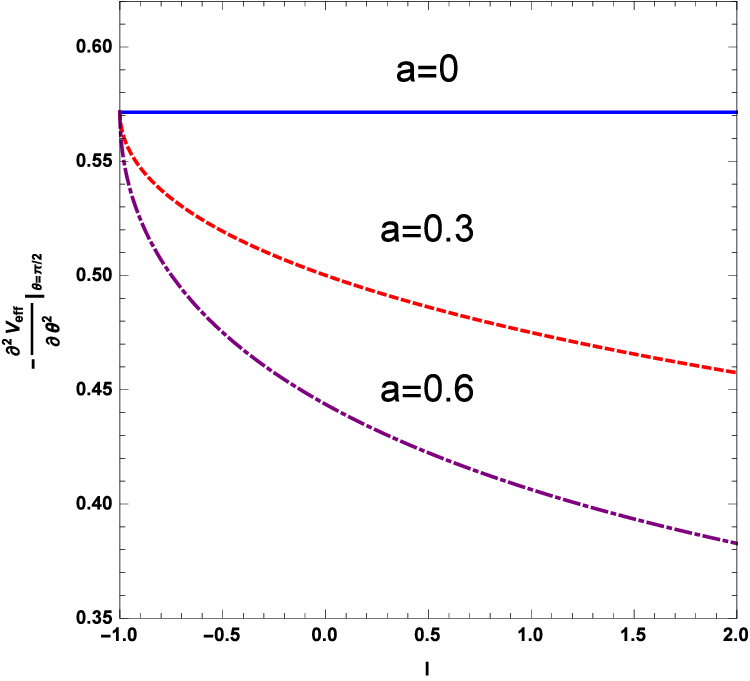} \includegraphics[width=5.5cm ]{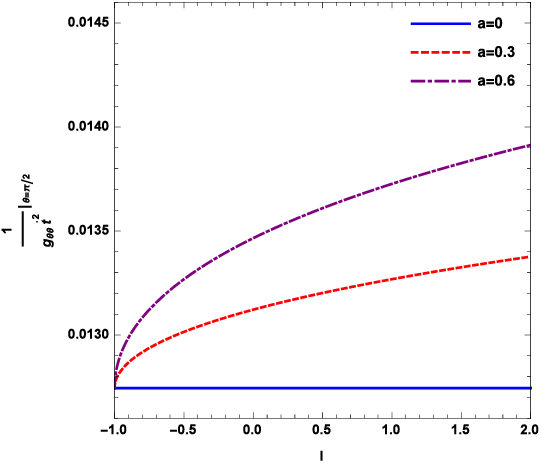}\includegraphics[width=5.0cm ]{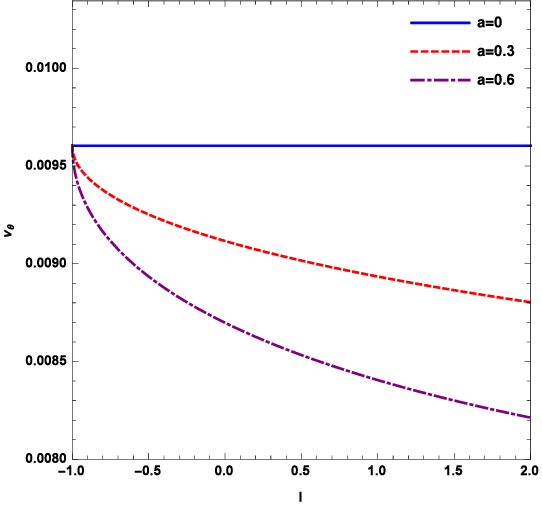}
\caption{The change of the second partial derivatives $-\frac{\partial^2 V_{eff}}{\partial r^2}|_{\theta=\frac{\pi}{2}}$, $-\frac{\partial^2 V_{eff}}{\partial \theta^2}|_{\theta=\frac{\pi}{2}}$, and the coefficients $\frac{1}{g_{rr}\dot{t}^2}|_{\theta=\frac{\pi}{2}}$, $\frac{1}{g_{\theta\theta}\dot{t}^2}|_{\theta=\frac{\pi}{2}}$ and  the frequencies $\nu_{r}$, $\nu_{\theta}$ with the parameter $l$ in the rotating black hole spacetime in Einstein-bumblebee theory. Here we set $M=1$ and $r=6.5$.}
\label{vddrr}
\end{figure}

Let us now focus on the stable circular orbits and assume some small perturbations around a stable circular orbit $r=r_0$ in the  equatorial plane \cite{TB1,TB101,TB2,TB3,TB4,TB5,TB6,TB7,TB8,TB9,TB10,TB11,TB12,TB13}, i.e.,
\begin{eqnarray}
r(t)=r_0+\delta r(t), \;\;\;\;\;\;\;\;\;\;\theta(t)=\frac{\pi}{2}+\delta \theta(t).
\end{eqnarray}
Inserting the above perturbations into Eq.(\ref{u3}), one can find that the perturbations $\delta r(t)$ and $\delta \theta(t)$ satisfy the following differential equations
\begin{eqnarray}
\frac{d^2\delta r(t) }{dt^2}+\Omega^2_{r}\delta r(t)=0, \;\;\;\;\;\;\;\;\;\;\frac{d^2\delta \theta(t) }{dt^2}+\Omega^2_{\theta}\;\delta \theta(t)=0,
\end{eqnarray}
with
\begin{eqnarray}
\Omega^2_{r}=-\frac{1}{2g_{rr}\dot{t}^2}\frac{\partial^2 V_{eff}}{\partial r^2}\bigg|_{r=r_0,\theta=\frac{\pi}{2}}, \;\;\;\;\;\;\;\;\;\;\Omega^2_{\theta}=-\frac{1}{2g_{\theta\theta}\dot{t}^2}\frac{\partial^2 V_{eff}}{\partial \theta^2}\bigg|_{r=r_0,\theta=\frac{\pi}{2}}.\label{jsdd0}
\end{eqnarray}
The radial epicyclic frequency $\nu_r$ and the vertical
epicyclic frequency $\nu_{\theta}$ can be written as $\nu_r=\Omega_r/2\pi$ and $\nu_{\theta}=\Omega_{\theta}/2\pi$, respectively. Inserting metric functions (\ref{metric}) into Eq.(\ref{jsd0}) ,
we can find the azimuthal frequency
\begin{eqnarray}\label{pinlv1}
\nu_{\phi}=\frac{1}{2\pi}\frac{M^{1/2}}{r^{3/2}_0+a^{*}M^{3/2}\sqrt{l+1}},
\end{eqnarray}
where $a^*\equiv a/M$. It is easy to find that the azimuthal frequency $\nu_{\phi}$ decreases with the Lorentz symmetry breaking parameter $l$ for the rotating case.  From  Eq.(\ref{jsd0}), one can find that this behavior of $\nu_{\phi}$ with $l$ is dominated by the derivatives $g_{\phi\phi,r}$ and $g_{t\phi,r}$ which increase with $l$ in the equatorial plane.
As $a=0$, one can find that $\nu_{\phi}$ is independent of the parameter $l$. Similarly, substituting metric functions (\ref{metric}) into Eqs.(\ref{t0}) and (\ref{jsdd0}), one has
\begin{eqnarray}
\nu_{r}&=&\nu_{\phi}\bigg[\frac{1}{l+1}-\frac{6M}{(l+1)r_0}+\frac{8a^{*}M^{3/2}}{\sqrt{l+1}r^{3/2}_0}
-3a^{*2}\frac{M^{2}}{r^{2}_0}
\bigg]^{1/2},\label{pinlv2}\\
\nu_{\theta}&=&\nu_{\phi}\bigg[1-\frac{4a^{*}\sqrt{l+1}M^{3/2}}{r^{3/2}_0}+3a^{*2}(l+1)
\frac{M^2}{r^{2}_0}\bigg]^{1/2}.\label{pinlv22}
\end{eqnarray}
Obviously, in the rotating case $a\neq 0$, the frequencies $\nu_{r}$ and $\nu_{\theta}$ depend on the Lorentz symmetry breaking parameter $l$.
However, in the non-rotating case, one can find that only the frequency $\nu_{r}$ is related to the parameter $l$  since  $\nu_{\theta}$ is identical with $\nu_{\phi}$ in this case  with $a=0$ and they are not functions of the parameter $l$.
The properties of above three frequencies make it possible to constrain effect from the Lorentz symmetry breaking  by quasi-periodic oscillations. As $l=0$, it is easy to find that these three frequencies reduce to those in the usual Kerr black hole spacetime \cite{RPM1,RPM2,RPM20,RPM3}. It is well known that the effective potential (\ref{veffpo}) plays an important role in determining the circular orbit's radius of particle and the corresponding
frequencies of motions. From Eq.(\ref{jsdd0}), the frequencies $\nu_{r}$ and $\nu_{\theta}$ are determined by the second derivatives of the effective potential (\ref{veffpo}) together with a factor related to metric function and $\dot{t}^2$. In Fig.(\ref{vddrr}), we show the change of the partial derivatives $\frac{\partial^2 V_{eff}}{\partial r^2}|_{\theta=\frac{\pi}{2}}$, $\frac{\partial^2 V_{eff}}{\partial \theta^2}|_{\theta=\frac{\pi}{2}}$, and the factors $\frac{1}{g_{rr}\dot{t}^2}|_{\theta=\frac{\pi}{2}}$, $\frac{1}{g_{\theta\theta}\dot{t}^2}|_{\theta=\frac{\pi}{2}}$ with $l$ for fixed $r_0=6.5$. It is shown that the absolute value of $\frac{\partial^2 V_{eff}}{\partial r^2}|_{\theta=\frac{\pi}{2}}$ increases with $l$, but the factor $\frac{1}{g_{rr}\dot{t}^2}|_{\theta=\frac{\pi}{2}}$ decreases. However, the effect of the second derivative $\frac{\partial^2 V_{eff}}{\partial r^2}|_{\theta=\frac{\pi}{2}}$ is suppressed by the factor $\frac{1}{g_{rr}\dot{t}^2}|_{\theta=\frac{\pi}{2}}$, which leads to that the frequency $\nu_{r}$ decreases with the parameter $l$. Since both the absolute value of $\frac{\partial^2 V_{eff}}{\partial r^2}|_{\theta=\frac{\pi}{2}}$ and $\frac{1}{g_{rr}\dot{t}^2}|_{\theta=\frac{\pi}{2}}$ increase with $a$, it is easy to obtain that the frequencies $\nu_{r}$ increases with $a$. Moreover, from Fig.(\ref{vddrr}),  we also find that the second derivative $\frac{\partial^2 V_{eff}}{\partial \theta^2}|_{\theta=\frac{\pi}{2}}$  dominates the change of frequency $\nu_{\theta}$ and results in that $\nu_{\theta}$ is a decreasing function of $l$ and $a$. Furthermore, the periastron and nodal precession frequencies can be expressed as
\begin{eqnarray}
\nu_{\text{per}}=\nu_{\phi}-\nu_{r},\;\;\;\;\;\;\;\;\;\;\;
\nu_{\text{nod}}=\nu_{\phi}-\nu_{\theta},
\end{eqnarray}
respectively.
\begin{figure}
\includegraphics[width=5.5cm ]{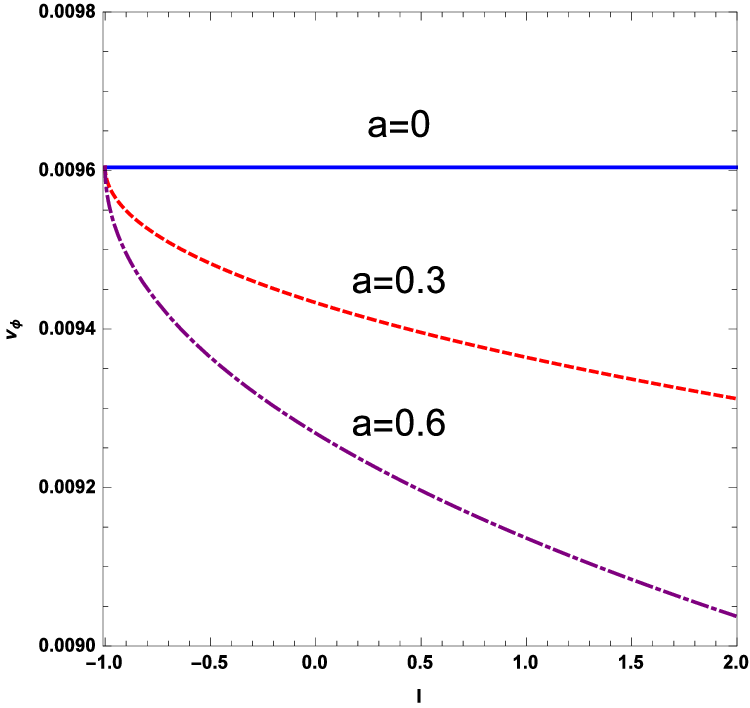} \includegraphics[width=5.5cm ]{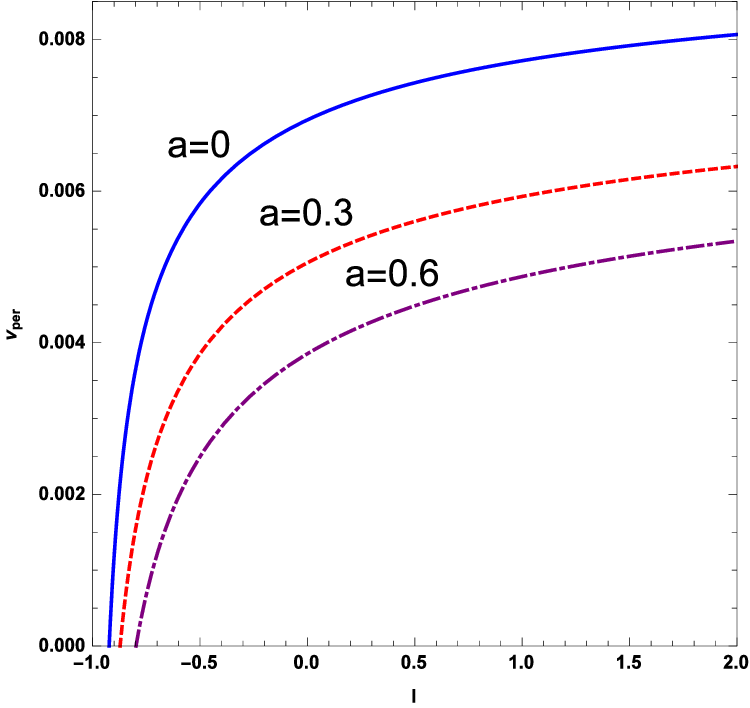}\includegraphics[width=5.5cm ]{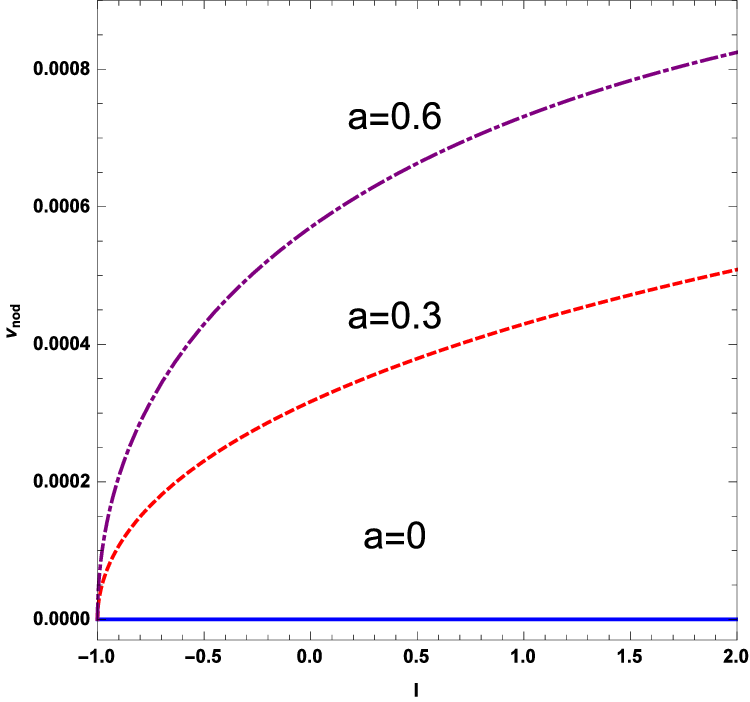}
\caption{The change of the frequencies $\nu_{\phi}$, $\nu_{per}$ and $\nu_{nod}$ with the parameter $l$ in the rotating black hole spacetime in Einstein-bumblebee theory. Here we set $M=1$ and $r=6.5$.}
\label{figure1}
\end{figure}
In Fig.(\ref{figure1}), we plot  the change of the frequencies $\nu_{\phi}$, $\nu_{per}$ and $\nu_{nod}$ for the rotating black hole spacetime in Einstein-bumblebee theory (\ref{metric}). It is shown that in the case with $a\neq0$ the azimuthal frequency $\nu_{\phi}$ decreases with $l$ as in the previous discussion. Comparing  Fig.(\ref{vddrr}) with Fig.(\ref{figure1}), one can find that the frequencies $\nu_{r}$ and $\nu_{\theta}$ decrease more rapidly than $\nu_{\phi}$, which yields that both of the periastron  and nodal precession frequencies ( $\nu_{per}$ and $\nu_{nod}$ ) increase with the Lorentz symmetry breaking parameter $l$. Thus, the changes of  $\nu_{per}$ and $\nu_{nod}$ with $l$ are determined by the effective potential combined with the factor related to metric function and $\dot{t}^2$.
We also find that as $a=0$ the nodal precession frequency $\nu_{nod}$ is zero for arbitrary $l$ as expected. With the increase of the spin parameter $a$, the frequencies $\nu_{\phi}$ and $\nu_{per}$ decrease, but the frequency $\nu_{nod}$ increases.

According to the relativistic precession model, three simultaneous quasi-periodic oscillations frequencies are generated at the same radius of the orbit in the accretion flow. For a rotating black hole spacetime  (\ref{metric}) in Einstein-bumblebee gravity, there are  three parameters to describe black hole spactime. Thus, we have to resort to the $\chi^2$ analysis and fit the values of these variables. Here, we adopt the observed data from black hole sources exhibiting high frequency quasi-periodic oscillations, which are listed in Table I. From the current observations of GRO J1655-40, there are two set of data about these frequencies ($\nu_{\phi}, \nu_{\text{per}},\nu_{\text{nod}}$ )\cite{RPM1,TB1}.
\begin{table}
\begin{tabular}{|c|c|c|c|c|}
\hline
&&&&\\
 \text{ } & \quad\quad\quad$\nu_{\phi}$ \quad\quad\quad & \quad\quad\quad $\nu_{\text{per}}$ \quad\quad\quad& \quad\quad\quad $\nu_{\text{nod}}$ \quad\quad\quad &\quad\quad\quad $M/M_{\odot}$\quad\quad\quad\\
\hline \multirow{5}{*}{\text { GRO J1655-40}} &&&&\\
& $441\pm2$\cite{RPM1} &$298\pm4$ \cite{RPM1} &$17.3\pm0.1$\cite{RPM1} & \multirow{5}{*}{$5.4\pm0.3$ \cite{TB4}}\\
&&&&\\
\cline{2-4}&&&&\\
&$451\pm5$ \cite{RPM1}&---&$18.3\pm0.1$ \cite{RPM1}&\\&&&&\\
\hline &&&&\\
\text { XTE J1550-564 } &$276\pm3$\cite{TB13}& $184\pm5$ \cite{TB13}& --- &$9.1\pm0.61$ \cite{XTE1}\\
&&&&\\
\hline &&&&\\
\text {GRS 1915+105} & $168\pm3$ \cite{TB13}& $113\pm5$ \cite{TB13}& --- &$12.4^{+2.0}_{-1.8}$ \cite{GRS1}\\&&&&\\
\hline
\end{tabular}
\caption{Data of quasi-periodic oscillations and black hole mass for GRO J1655-40, XTE J1550-564, and GRS 1915+105, respectively.}
\end{table}
Two set of frequencies can be regarded to be emitted by the relativistic particles moving along the orbits with the different radius $r_1$ and $r_2$, respectively. Moreover,  the mass of the black hole  is also independently measured by a dynamical method \cite{TB4}: $M_{\text{dyn}}=5.4\pm0.3M_{\odot}$. For the black hole sources XTE J1550-564  and GRS 1915+105, there are only the high frequencies data and the low frequency parts are lacking.
With the data listed in Table I, we can constrain the parameters of a rotating black hole spacetime  (\ref{metric}) in Einstein-bumblebee gravity through the relativistic precession model as in ref.\cite{TB1}. Together with the $\chi^2$ analysis,  we can fit the parameters of black hole (\ref{metric}) in Einstein-bumblebee gravity.
\begin{table}
\begin{tabular}{|c|c|c|c|c|}
\hline
&&&&\\
 \text{ } & \quad\quad\quad$M/M_{\odot}$ \quad\quad\quad & \quad\quad\quad $
 a^{*}$ \quad\quad\quad& \quad\quad\quad $l$ \quad\quad\quad &\quad\quad\quad $r/M$\quad\quad\quad \quad\quad\quad\\
\hline  &&&&\\ &&&&$r_1=5.6194^{+0.0346}_{-0.0334}$\\
\text { GRO J1655-40}
& $5.4002^{+0.0478}_{-0.0562}$ &$ 0.2976^{+0.0233}_{-0.0119}$ & $-0.1048^{+0.1678}_{-0.1316}$&\\
&&&&$r_2=5.5154^{+0.0476}_{-0.0474}$\\
&&&&\\
\hline &&&&\\
\text { XTE J1550-564 } &$9.100^{+0.2450}_{-1.1443}$ & $ 0.3697^{+0.4536}_{-0.0436}$ &$-0.2053^{+6.7573}_{-0.3635} $ &$5.4030^{+0.1010}_{-0.4050} $\\
&&&&\\
\hline &&&&\\
\text {GRS 1915+105} & $12.4000^{+0.7400}_{-3.3580}$ & $0.3080^{+3.7760}_{-0.3192}$ & $1.3083^{+9.5717}_{-2.0134}$ &$6.101^{+0.2566}_{-1.4794}$\\&&&&\\
\hline
\end{tabular}
\caption{Best-fit values and their range of $1\sigma$ for the black hole parameters with the metric (\ref{metric}) from  GRO J1655-40, XTE J1550-564, and GRS 1915+105, respectively.}
\end{table}
The best-fit values and their range of $1\sigma$ for the black hole parameters are listed in Table II.
\begin{figure}
\includegraphics[width=5cm ]{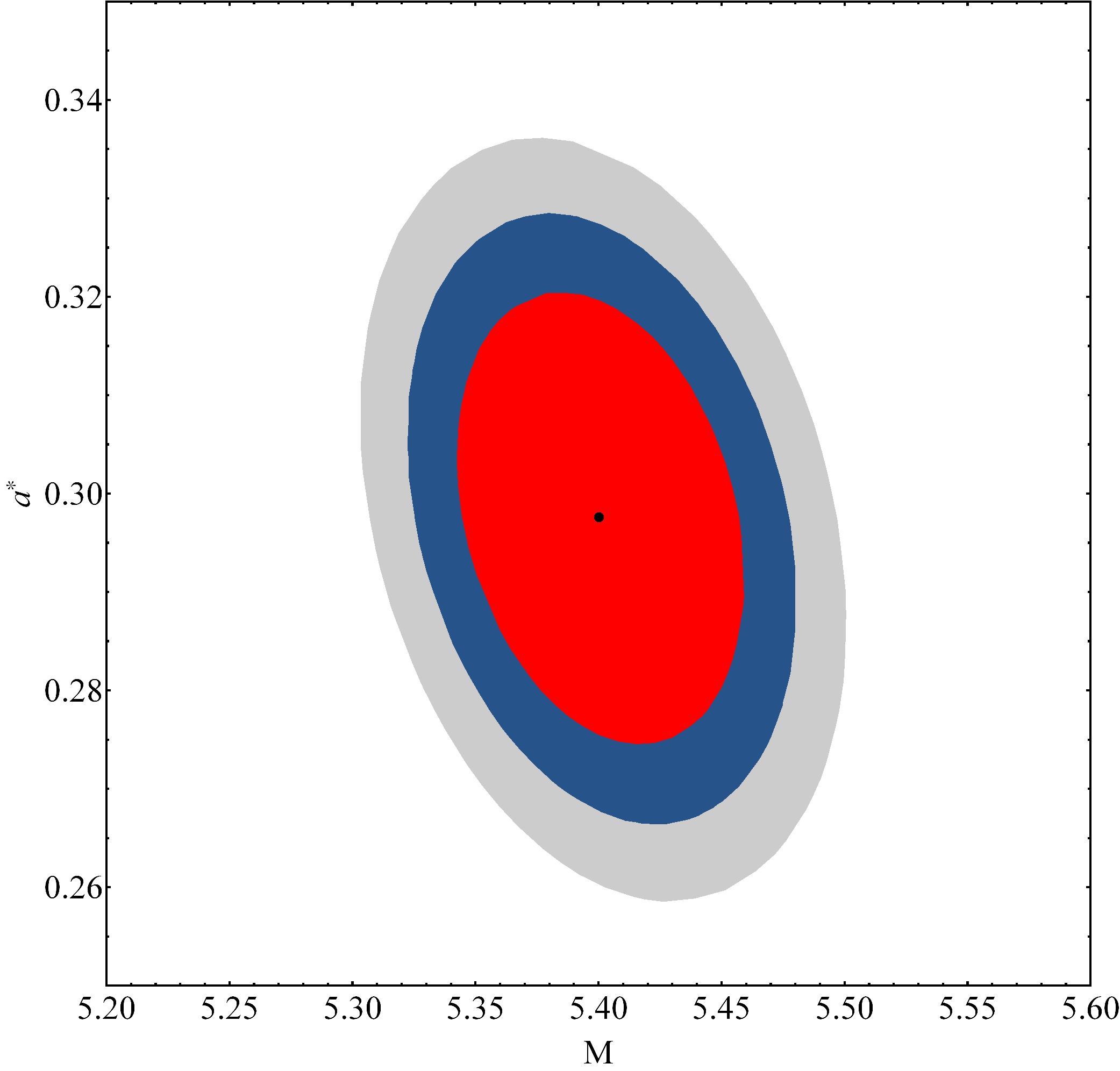}\;\; \includegraphics[width=5cm ]{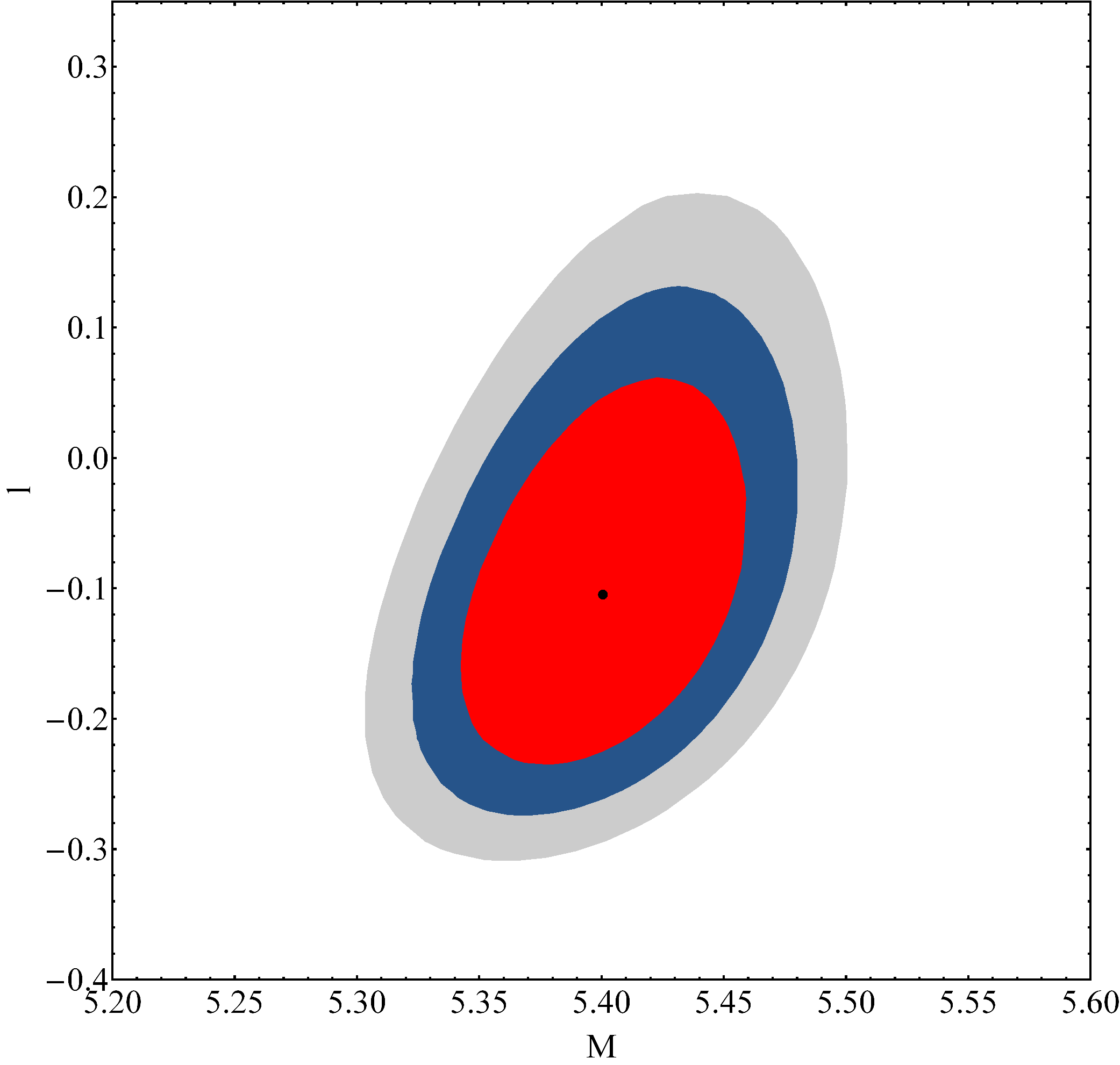} \;\; \includegraphics[width=5cm ]{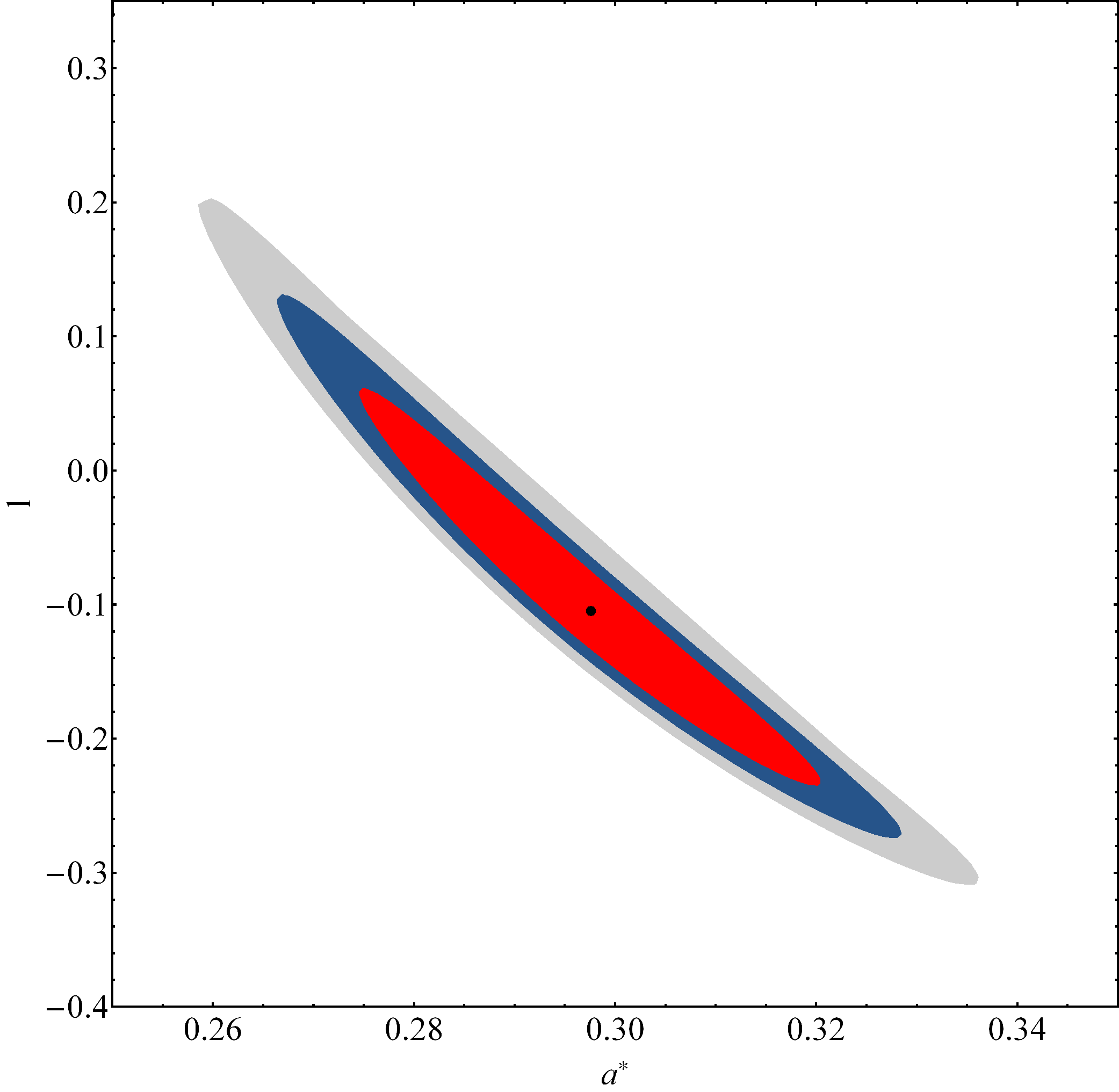}\\
\includegraphics[width=5cm ]{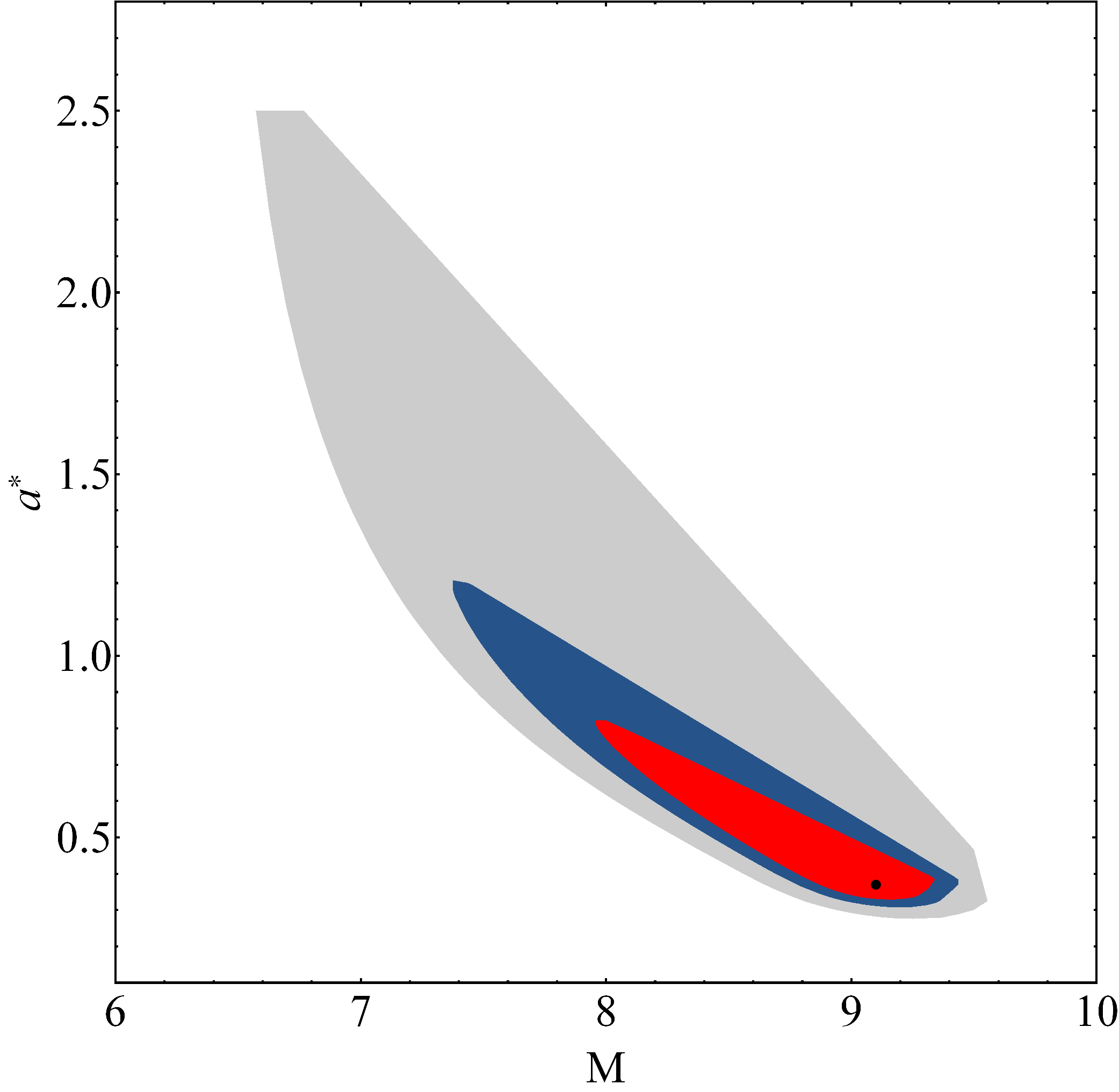} \;\;\includegraphics[width=5cm ]{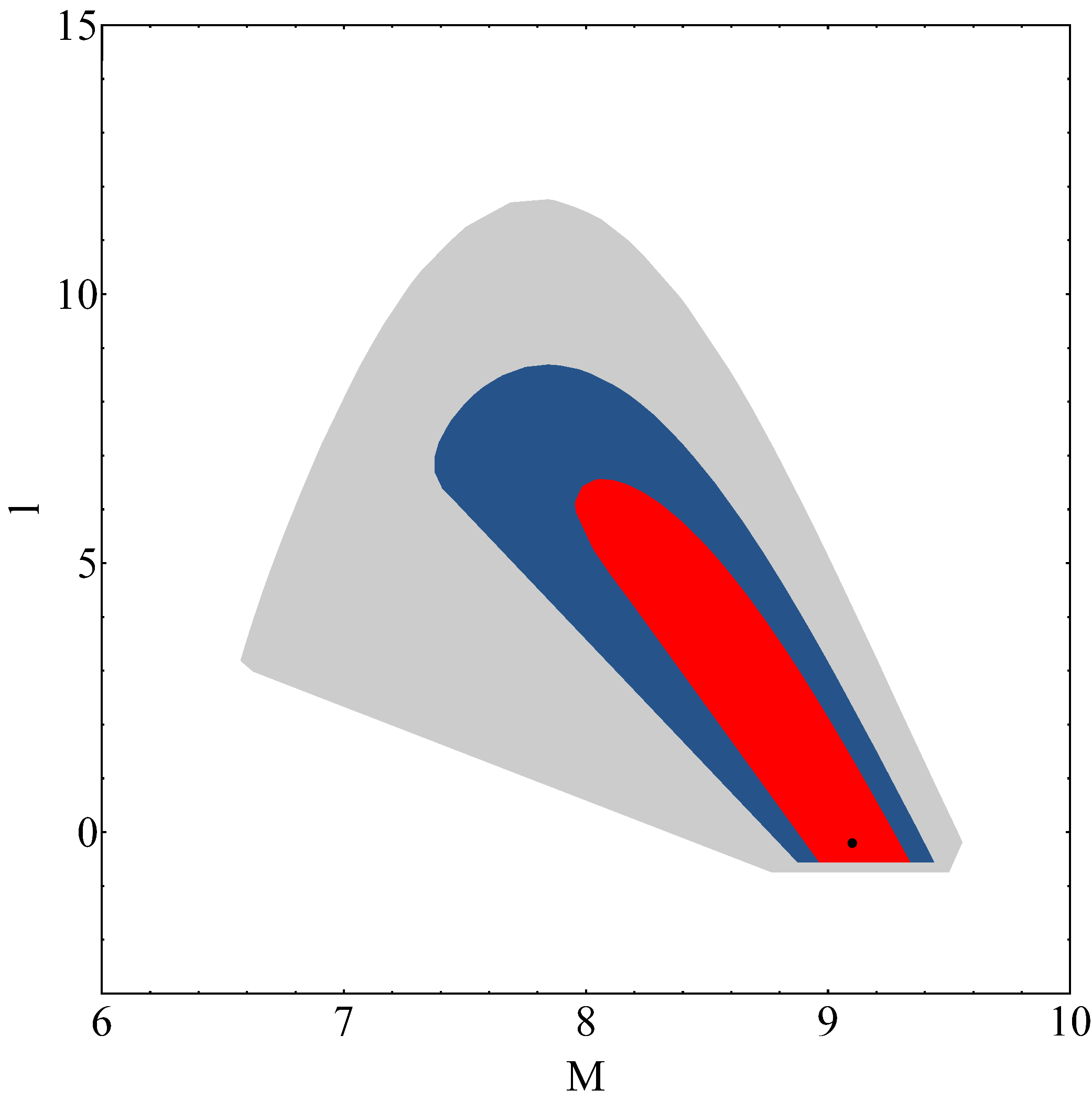} \;\; \includegraphics[width=5cm ]{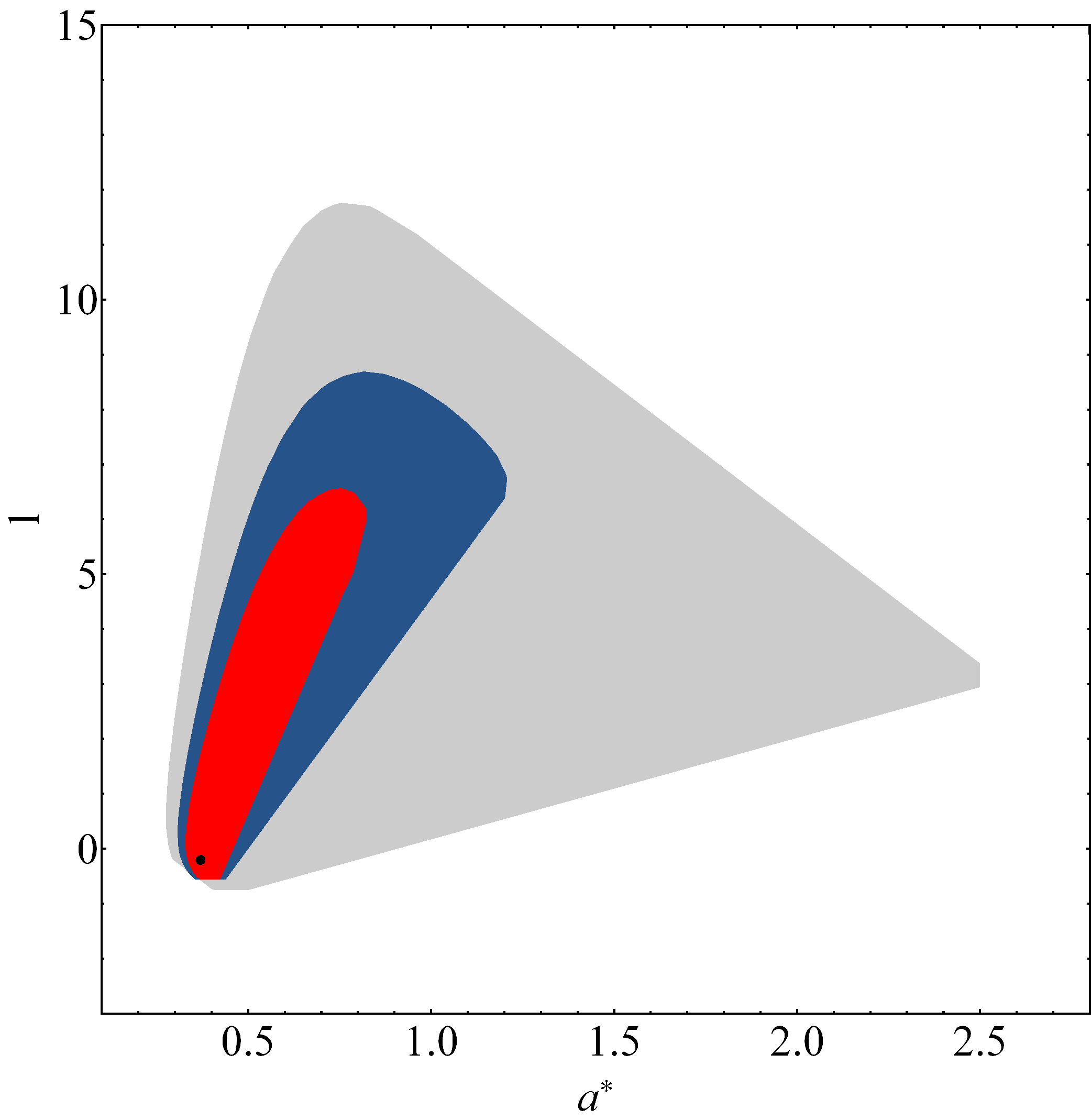}\\
\includegraphics[width=5cm ]{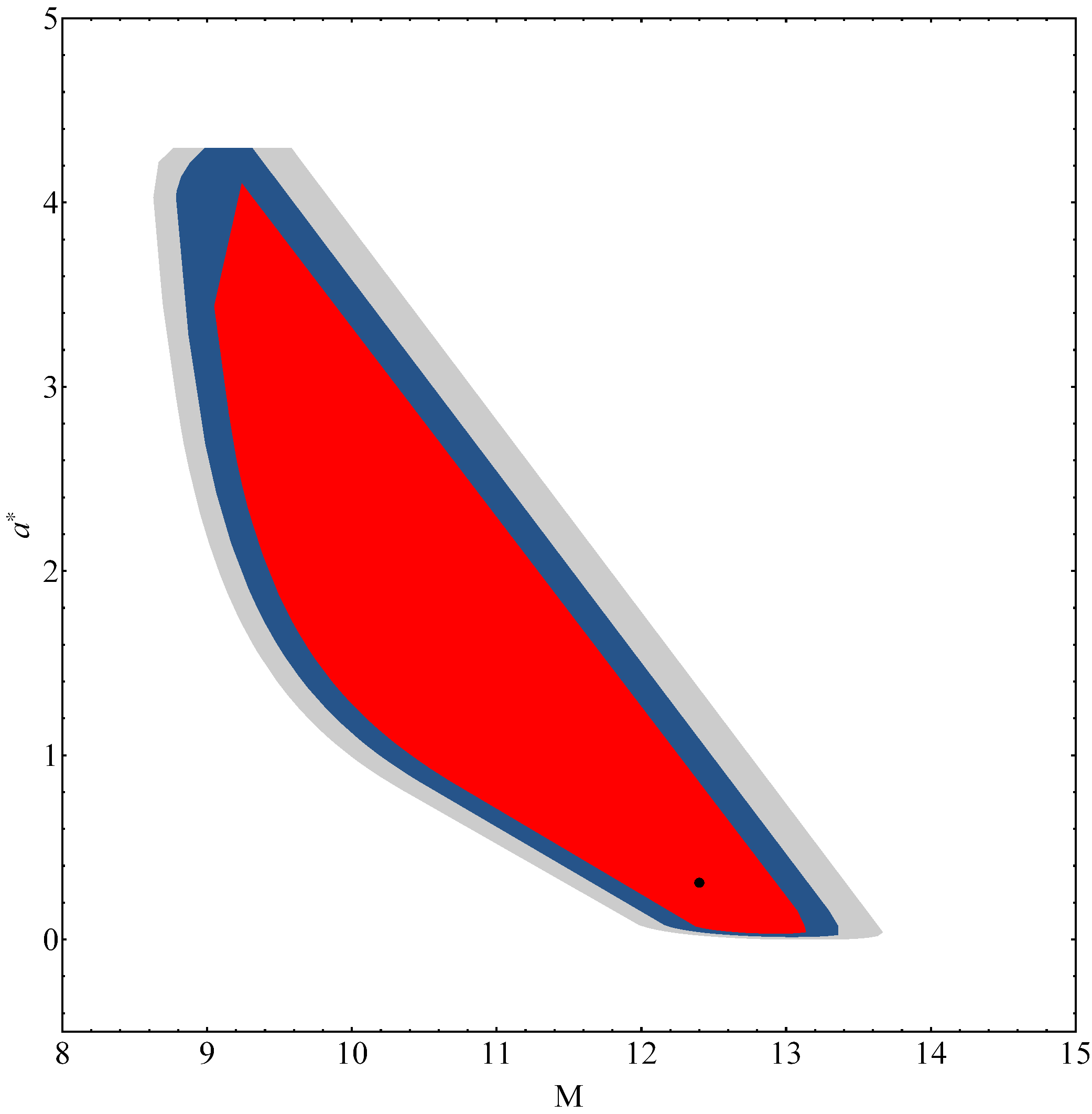} \;\;\includegraphics[width=5cm ]{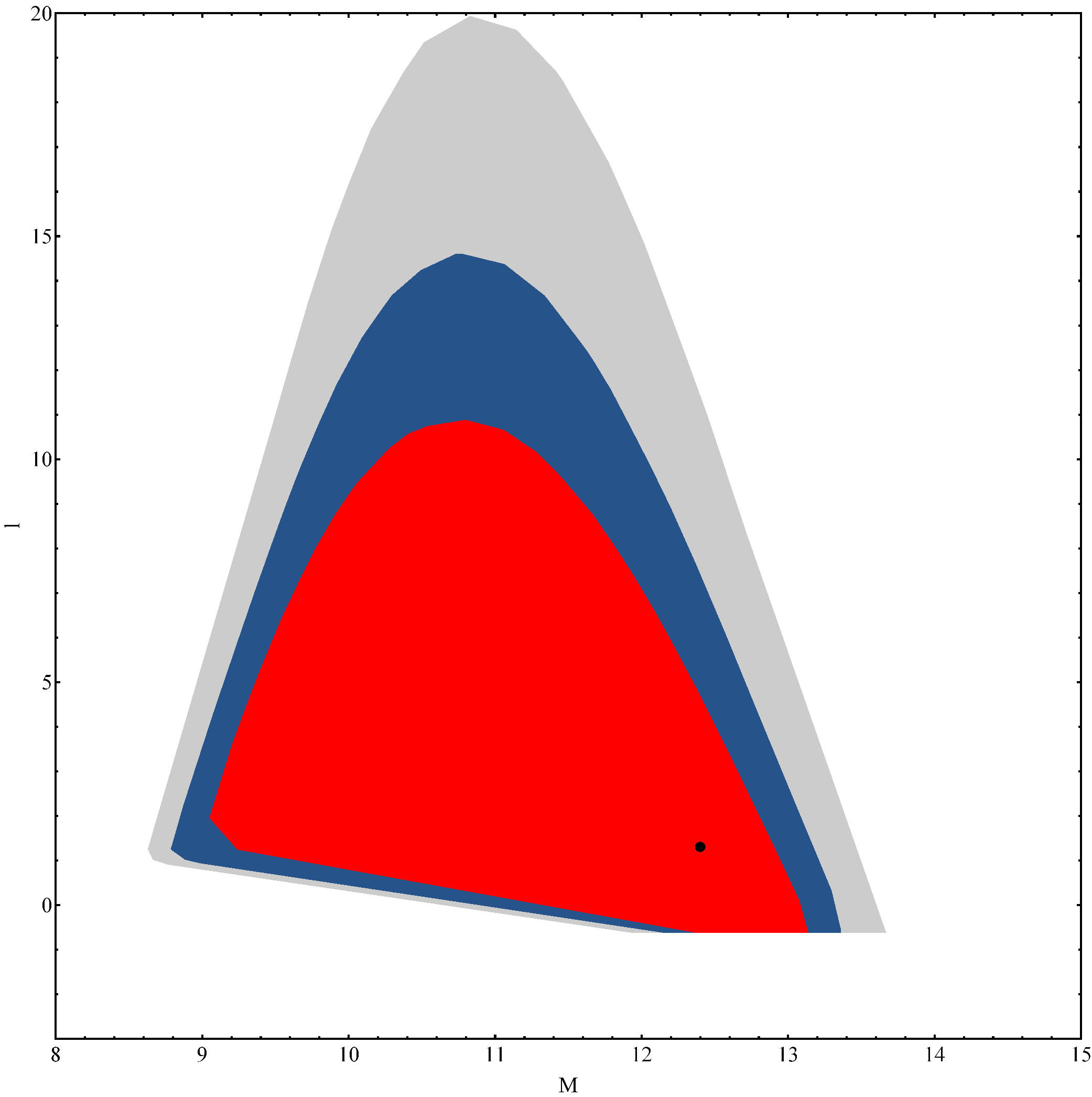} \;\; \includegraphics[width=5cm ]{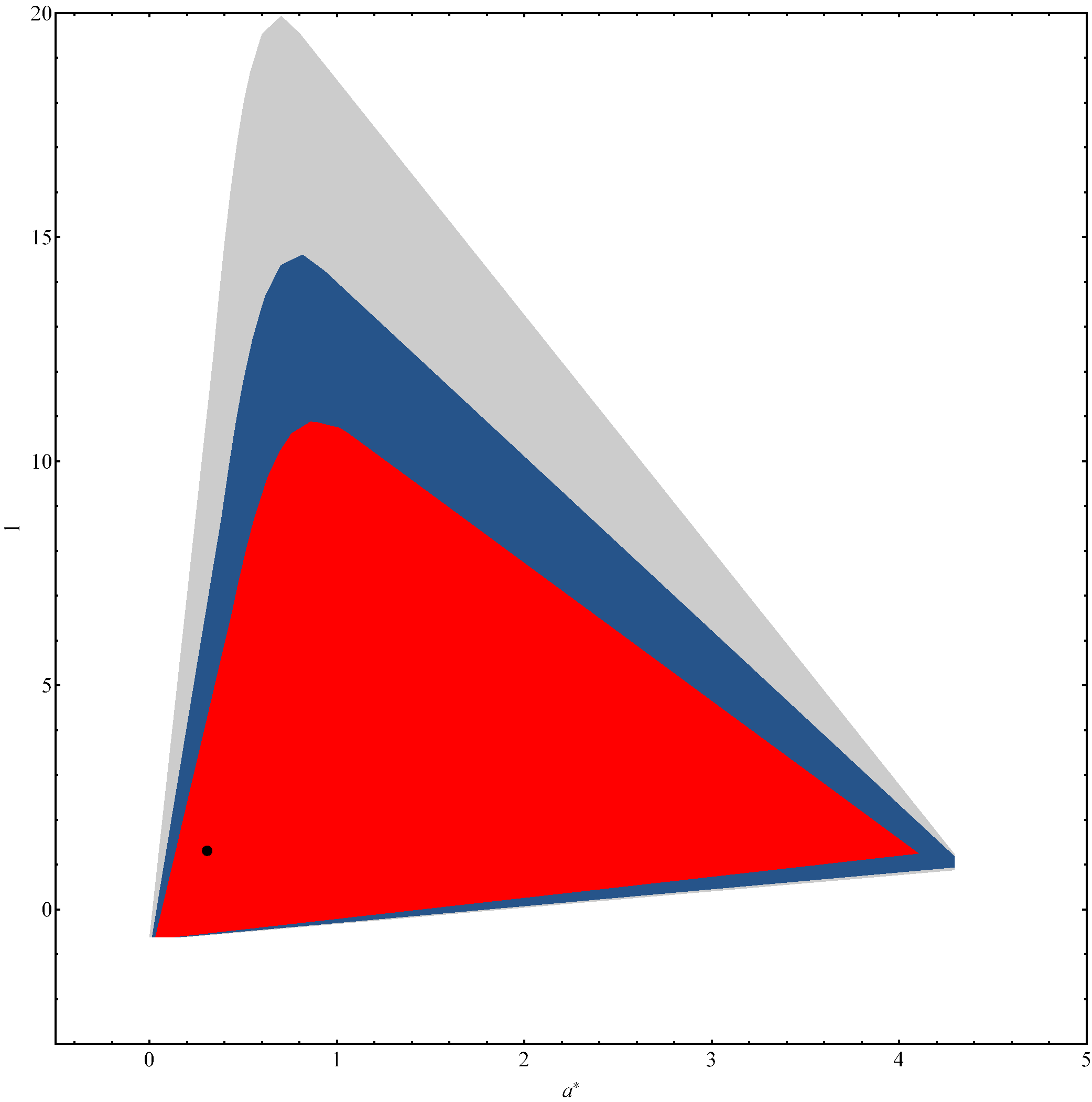}
\caption{Constraints on the parameters of the rotating black hole in Einstein-bumblebee theory (\ref{metric}) from current observations of QPOs within the relativistic precession model. The top, middle and bottom rows correspond to the constraint from  GRO J1655-40, XTE J1550-564, and GRS 1915+105, respectively.
The red, blue and gray regions in the panels represent the contour levels $1\sigma$, $2\sigma$ and $3\sigma$, respectively. The black dots denote the best-fit values of black hole parameters.}
\label{figure2bb}
\end{figure}
The Table II shows that the circular orbit of quasi-periodic oscillations lies in the strong gravitational-field region of the black hole. In Fig.\ref{figure2bb}, we show the contour levels of $1\sigma$, $2\sigma$ and $3\sigma$ for the black hole parameters $M$, $a$ and $l$ with different observed black hole sources. Comparing with the constraint results obtained by data of three black hole sources, presented in Table II and Fig.\ref{figure2bb}, we find  that the $1\sigma$ region of $l$ obtained by GRO J1655-40 data is the most narrow and it also lies in the  $1\sigma$ regions obtained by the other two black hole sources, which means that
 the constraint on the Lorentz symmetry breaking parameter $l$ is more precise with data of GRO J1655-40.
The main reason may be that there are more available observation data of quasi-periodic oscillations for GRO J1655-40.

According to the constraint from  GRO J1655-40, the best-fit value of $l=-0.1048$  is negative, which means that the spacetime described gravitational field in the Einstein-bumblebee gravity (\ref{metric}) should allow $|a|/M>1$ for a black hole. It implies that the range of black hole spin parameter $a$ is larger than that in the Kerr case in general relativity. Comparing with the usual Kerr black hole spacetime, the negative $l$ leads to that both the outer ergosurface radius $r_{outerg}$ and the outer horizon radius $r_+$ increase, but the width between the outer ergosurface and the outer horizon $r_{outerg}-r_+=\frac{a^2\sin^2\theta}{\sqrt{M^2-(l+1)a^2}+\sqrt{M^2-(l+1)a^2\cos^2\theta}}$ decreases
for fixed $\theta$, which yields the lower possibility of exacting energy by Penrose process for a rotating black hole in Einstein-bumblebee gravity (\ref{metric}). Moreover, the negative $l$ means that the black hole (\ref{metric}) owns the higher Hawking temperature and the stronger Hawking radiation than the Kerr black hole. From Table II and Fig.(2), we find that the case of $l = 0$ lies in the range of $1\sigma$ obtained by three black hole sources, which means that general relativity remains to be consistent with the observation data of quasi-periodic oscillations frequencies.

\section{summary}

With relativistic precession model, we have studied  quasi-periodic oscillations frequencies in a rotating black hole in Einstein-bumblebee gravity (\ref{metric}). The black hole owns three parameters: mass $M$, spin $a$ and the Lorentz symmetry breaking parameter $l$. We find that in the case with $a\neq0$ both of the periastron  and nodal precession frequencies ( $\nu_{per}$ and $\nu_{nod}$ ) increase with the Lorentz symmetry breaking parameter $l$, but the azimuthal frequency $\nu_{\phi}$ decreases. In the non-rotating black hole case, the nodal precession frequency $\nu_{nod}$ is zero for arbitrary $l$ since $\nu_{\theta}=\nu_{\phi}$ in this case and they are independent of the parameter $l$.  With the increase of the spin parameter, the frequencies $\nu_{\phi}$ and $\nu_{per}$ decrease, but the frequency $\nu_{nod}$ increases. With the observation data of GRO J1655-40, XTE J1550-564, and GRS 1915+105,  we constrain the parameters of the rotating black hole in Einstein-bumblebee gravity (\ref{metric}), respectively.  Our results show that the constraint on the Lorentz symmetry breaking parameter $l$ is more precise with data of GRO J1655-40. According to the constraint from  GRO J1655-40, one can find that the best-fit value of the Lorentz symmetry breaking parameter $l$ is negative. Comparing with the usual Kerr spacetime, the negative $l$ leads to that the black hole (\ref{metric}) in Einstein-bumblebee gravity owns the higher Hawking temperature and the stronger Hawking radiation than the Kerr black hole, but the lower possibility of exacting energy by Penrose process.
However, in the range of $1 \sigma$, general relativity (where$l = 0$) remains to be consistent with the observation data of GRO J1655-40, XTE J1550-564 and GRS 1915+105.

\section{\bf Acknowledgments}
This work was  supported by the National Natural Science
Foundation of China under Grant No.11875026, 11875025, 12035005 and 2020YFC2201403.

\section{appendix}
In this section, we present the derivation of the equation (\ref{r0gdao}) for the circular orbit $r_0$. The effective potential (\ref{vpequator}) can be written as
\begin{eqnarray}
V_{eff}\equiv\frac{A(r)}{B(r)}-1,
\end{eqnarray}
with
\begin{eqnarray}
A(r)&\equiv & [r^3+(r+2M)(l+1)a^2]E^2-4aM\sqrt{l+1}EL_z-(r-2M)L^2_z,\nonumber\\
B(r)&\equiv & r[r^2-2Mr+(l+1)a^2].
\end{eqnarray}
From the conditions (\ref{vcondition}) of the circular orbit, one can obtain
\begin{eqnarray}\label{conqq3}
A(r_0)=B(r_0), \quad\quad\quad A(r_0)B'(r_0)-A'(r_0)B(r_0)=0.
\end{eqnarray}
It means that $A'(r_0)=B'(r_0)$, which gives directly the equation (\ref{L0gdao}). Substituting it into the above equations (\ref{conqq3}), one can get the equation (\ref{r0gdao}) satisfied by the circular orbit $r_0$ in the equatorial plane.

\end{document}